\documentclass[%
 reprint,
superscriptaddress,  % <<========
 amsmath,amssymb,
 aps,
]{revtex4-1}

\usepackage{graphicx}% Include figure files
\usepackage{dcolumn}% Align table columns on decimal point
\usepackage{bm}% bold math

\begin{document}

\preprint{APS/123-QED}

\title{Origin of magnetic moments and ferromagnetic properties of\\ potassium clusters in zeolite A}% Force line breaks with \\

\author{Takehito~Nakano}
\affiliation{Institute of Quantum Beam Science, Graduate School of Science and Engineering, Ibaraki University, \\2-1-1 Bunkyo, Mito, Ibaraki 310-8512, Japan}
\affiliation{Department of Physics, Graduate School of Science, Osaka University, \\1-1 Machikaneyama, Toyonaka, Osaka 560-0043, Japan}
\author{Shingo~Araki}
\affiliation{Department of Physics, Okayama University, Okayama 700-8530, Japan}
\author{Nguyen~Hoang~Nam}
\affiliation{Center for Materials Science, Faculty of Physics, Hanoi University of Science, Vietnam National University, \\334 Nguyen Trai, Thanh Xuan, Hanoi, Vietnam}
\author{Takashi~Umemoto}
\affiliation{Department of Physics, Graduate School of Science, Osaka University, \\1-1 Machikaneyama, Toyonaka, Osaka 560-0043, Japan}
\author{Kazushige~Tsuchihashi}
\affiliation{Department of Physics, Graduate School of Science, Osaka University, \\1-1 Machikaneyama, Toyonaka, Osaka 560-0043, Japan}
\author{Yosuke~Kubo}
\affiliation{Department of Physics, Graduate School of Science, Osaka University, \\1-1 Machikaneyama, Toyonaka, Osaka 560-0043, Japan}
\author{Akihiro~Owaki}
\affiliation{Department of Physics, Graduate School of Science, Osaka University, \\1-1 Machikaneyama, Toyonaka, Osaka 560-0043, Japan}
\author{Yasuo~Nozue}
\email{nozue@phys.sci.osaka-u.ac.jp}
\affiliation{Department of Physics, Graduate School of Science, Osaka University, \\1-1 Machikaneyama, Toyonaka, Osaka 560-0043, Japan}

%\affiliation{}
%\email{araki@science.okayama-u.ac.jp}

%\author{Nguyen Hoang Nam}
%\affiliation{%Center for Materials Science, Faculty of Physics, Hanoi University of Science, Vietnam National University,\\334 Nguyen Trai, Thanh Xuan, Hanoi, Vietnam}

%\author{Kota Shimodo}
%\affiliation{%
%Department of Physics, Graduate School of Science, Osaka University, \\1-1 Machikaneyama, Toyonaka, Osaka 560-0043, Japan}


%\affiliation{
%\affiliation{
%\affiliation{%


\date{\today}% It is always \today, today,

\begin{abstract}
K clusters arrayed in zeolite A are investigated in detail. K clusters are generated in regular $\alpha$-cages of zeolite A by the loading of guest K metal at a loading density of K atoms per $\alpha$-cage, $n$.  The value of $n$ was changed from 0 to 7.2. It is known that this system shows ferromagnetic properties for $n > 2$, where the Curie temperature increases with $n$, has a peak of $\approx\,$8 K at $n \approx 3.6$, and decreases to 0 K at $n = 7.2$. The negative Weiss temperature is estimated from the Curie-Weiss law for $n > 2$. A spherical quantum-well (SQW) model for the K cluster with 1$s$, 1$p$, and 1$d$ quantum states has previously been proposed, and the ferromagnetic properties were explained as being due to $s$-electrons in 1$p$ states. A spin-cant model of Mott-insulator antiferromagnetism in a K cluster array has been proposed for the origin of the ferromagnetic properties.  However, the SQW model cannot explain either the $n$-dependence of the Curie temperature or that of the Curie constant. In the present study, detailed measurements were made of the optical reflection, magnetic properties, electron spin resonance, and electrical resistivities.  An optical reflection band at 0.7 eV is clearly observed for $2 < n < 6$ at low temperatures, and is assigned to the excitation of the $\sigma$-bonding state between 1$p$-states in adjoining $\alpha$-cages. The electrical resistivity indicates an insulating state continuously for $n$.  We propose an advanced SQW model with consideration for $\sigma$-bonding, the orbital orthogonality of 1$p$-hole states, and also the superlattice structure. We explain the observed electronic properties using this model based on a correlated polaron system. We propose an enhancement effect of the Dzyaloshinsky-Moriya (DM) interaction  between adjoining $1p$ orbitals by the extension of the Rashba mechanism of spin-orbit interaction. 
\begin{description}
\item[PACS numbers] 82.75.Vx, 75.50.Xx, 75.75.-c, 36.40.-c, 71.70.Ej, 71.28.+d
%75.75.+a, 75.50.Cc, 75.10.Lp, 73.22.-f, 82.75.Vx
\end{description}
\end{abstract}

\maketitle

%\tableofcontents

%\section{\label{sec:level1}First-level heading:\protect\\ The line break was forced \lowercase{via} \textbackslash\textbackslash}

%%%%%%%%%%%%%%%%%%%%%%%%%%%%%%%%%%%%%
\section{\label{sec:level1}Introduction}
Pauli paramagnetism of $s$-electrons in bulk alkali metals is quite different from ferromagnetism or antiferromagnetism of $d$-electrons with localized magnetic moments in transition metals.  Alkali metals can be loaded into regular nanospaces (cages) of zeolite crystals, and the $s$-electrons are localized in cages by the formation of clusters. These localized $s$-electrons of clusters interact with those in adjacent cages through windows of the cages and with those in the same cages.  The electronic properties of alkali metals in zeolites, such as ferrimagnetism, ferromagnetism, antiferromagnetism, and insulator-to-metal transitions, have been studied intensively~\cite{Nakano2017-APX, Nakano2013ICMInv}.   

The first discovery of ferromagnetic properties of alkali-metal clusters was made in potassium loaded zeolite A~\cite{Nozue1992-KA, Nozue1993-KA}, where $\alpha$-cages with the inside diameter of $\approx\,$11 \AA{} are arrayed in a simple cubic structure.   After detailed studies~\cite{Nakano2017-APX, Nakano2013ICMInv, Kodaira1993-KA, Nozue1993-KA, Armstrong1994, Nakano1999-KA, Maniwa1999, Ikeda2000KA, Nakano2000MCLC, Nakano2000, Nakano2001-RbA-KA, Kira2001, Kira2002, Nakano2002ESR, Arita2004, Aoki2004, Ikeda2004KA, Nakano2004, Nakano2007, Nam2007, Nakano2007-HighMag, Nakano2009muSR-KA, Nohara2009, Nohara2011}, a spin-cant model of Mott-insulator antiferromagnetism of K cluster array in $\alpha$-cages was proposed \cite{Nakano2017-APX, Nakano2004, Nakano2007}.  However, there is still uncertainty around the origin of the magnetic moments of K clusters in zeolite A.  The present paper details various extensive studies and new insights into the origin of the magnetic moments of K clusters in accordance with the latest theoretical calculations.

Zeolite crystals have free spaces in different regular cages for guest materials~\cite{Nakano2017-APX}. There are many different types of zeolite structures \cite{IZA}. Besides K clusters in zeolite A, various physical properties have been studied in different guest materials in different zeolites.  In Rb clusters in zeolite A, a ferrimagnetism has been observed~\cite{Nakano2001-RbA-KA, Duan2007, Duan2007b}.   In sodalite, $\beta$-cages with the inside diameter of $\approx\,$7 \AA{} are arrayed in a body centered cubic structure. An antiferromagnetism of Mott insulator has been clearly observed~\cite{Srdanov1998}, and detailed studies have been made~\cite{Sankey1998, Blake1998, Blake1999, Heinmaa2000, Madsen2001, Tou2001, Scheuermann2002, Madsen2004, Nakamura2009, Nakano2010, Nakano2012-SOD, Nakano2013JKPS, Nakano2013PRB, Nakano2015Mossbauer}.  In zeolite low-silica X (LSX), supercages with the inside diameter of $\approx\,$13 \AA{} and $\beta$-cages are arrayed in a diamond structure, respectively, namely the double diamond structure.  In Na-K alloy clusters in zeolite LSX, a N\'eel's N-type ferrimagnetism, a ferromagnetism, an anomalous temperature dependence of paramagnetic susceptibility, an insulator-to-metal transition, etc. have been observed~\cite{Nakano2017-APX, Nakano2013ICMInv, Hanh2010, Nam2010, Nozue-Na-LSX2012, Nakano2013K-LSX, Igarashi2013-Na-LSX, Kien2015, Igarashi2016, Araki2019}.  Alkali metals in quasi-low-dimensional systems, such as the quasi-one-dimensional metallic system in channel-type zeolite L \cite{Kelly1995, Anderson1997, Thi2016, Thi2017}, has been studied.  Besides alkali metals, novel properties of superfluidity have been observed in quantum liquids of $^4$He~\cite{Wada2009, Yamashita2009, Eggel2011, Taniguchi2013, Matsushita2016} and $^3$He~\cite{Taniguchi2005}.  Clusters of semiconducting materials, such as Se, PbI$_2$, HgI$_2$, BiI$_3$, etc., are also incorporated into zeolites, and novel properties related to the quantum confinement of electrons in nanoscale have been observed \cite{Bogomolov1978, Nozue-Se-1990, Tang1991, Tang1992, Demkov1995, Demkov1996a, Demkov1996b, Demkov1997, Demkov2001, Goldbach2005}.

\begin{figure}[ht]
\resizebox*{7.0cm}{!}{\includegraphics{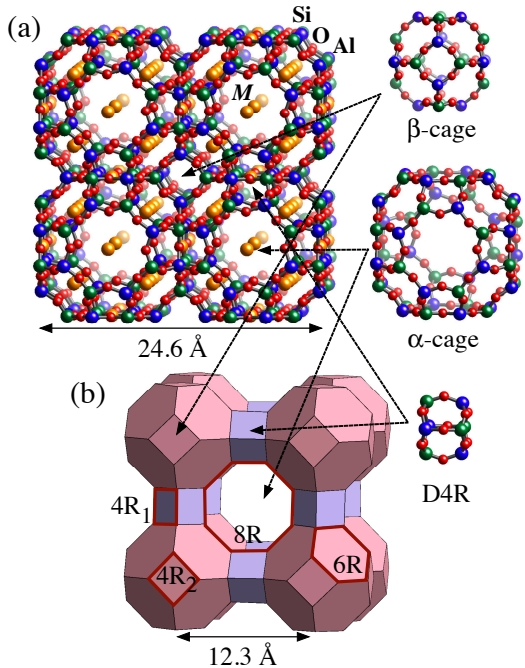}}
\caption{\label{fig:LTA}(Color online) (a) Aluminosilicate framework structure of zeolite A and typical sites of $M$ cations without guest materials. The space group is $Fm{\rm{\bar 3c}}$ and the lattice constant is 24.6 \AA, where adjoining $\alpha$-cages are crystallographically equivalent to each other. (b) Polyhedral illustration of LTA framework disregarding the alternate order of Al and Si atoms.  $\beta$-cages and $\alpha$-cages are arrayed in a simple cubic structure by the sharing of eight-membered rings (8Rs) and double-four-membered rings (D4Rs), respectively.}
\end{figure}
%%%%%%%%

\subsection{\label{sec:level2}Zeolite A}

Zeolite A is one of the most typical aluminosilicate zeolites, and is a nonmagnetic insulator unless it is loaded with guest materials. The negatively charged aluminosilicate framework of zeolite A is illustrated in Fig.~\ref{fig:LTA}(a) together with typical sites of exchangeable monovalent cations ($M$s).  Al and Si atoms are alternately connected by the sharing of O atoms. The space group is $Fm{\bar 3}c$ with a lattice constant of 24.6 \AA{}.  The chemical formula per unit cell is $M_{96}$Al$_{96}$Si$_{96}$O$_{384}$ before the loading of guest materials. The unit cell contains eight $\alpha$-cages (or eight $\beta$-cages), with adjoining $\alpha$-cages (or adjoining $\beta$-cages) being crystallographically equivalent to each other. 

If we disregard the alternate order of Al and Si atoms in the framework, the space group is $Pm{\bar 3}m$ with a lattice constant of 12.3 \AA{}.  The chemical formula per unit cell is $M_{12}$Al$_{12}$Si$_{12}$O$_{48}$.   This framework structure of zeolite A is called LTA (IUPAC nomenclature \cite{IZA}), and there is one $\alpha$-cage (or one $\beta$-cage) in the unit cell. A polyhedral illustration of LTA framework is given in Fig.~\ref{fig:LTA}(b) based around a $\beta$-cage.  The $\alpha$-cages and $\beta$-cages are arrayed in a simple cubic structure by the sharing of eight-membered rings (8Rs) and double-four-membered rings (D4Rs), respectively.  An $\alpha$-cage is constructed of six 8Rs, eight six-membered rings (6Rs), and twelve four-membered rings (4R$_1$s). A $\beta$-cage is constructed of eight 6Rs and six four-membered rings (4R$_2$s).  $\beta$-cages and $\alpha$-cages share 6Rs with each other. The effective inside diameters of an $\alpha$-cage and a $\beta$-cage are $\approx\,$11 and $\approx\,$7 \AA{}, respectively. The effective inside diameter of an 8R and a 6R are $\approx\,$5 and $\approx\,$3 \AA{}, respectively.  In the present study, we used K-form A, namely $M =$ K.  Hereafter, the present zeolite A is called K$_{12}$-A.

\begin{figure}[ht]
\resizebox*{7.2cm}{!}{\includegraphics{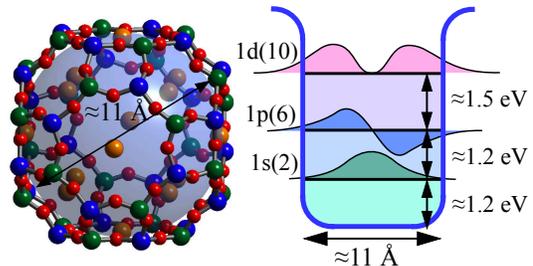}}
\caption{\label{fig:Cluster}(Color online) Schematic illustration of a K cluster in an $\alpha$-cage and the quantum states of an $s$-electron in a spherical quantum-well (SQW) potential with a diameter of $\approx$11 \AA. }
\end{figure}

\subsection{\label{sec:Cluster}Properties of K clusters in K$_n$/K$_{12}$-A}

By the loading of guest K metal into zeolite K$_{12}$-A, the 4$s$-electrons of the K atoms are shared with the K cations of the zeolite, and K clusters are formed in the $\alpha$-cages.  The chemical formula of K-loaded K$_{12}$-A is K$_{n}$$\cdot$K$_{12}$Al$_{12}$Si$_{12}$O$_{48}$, where $n$ is the average number of guest K atoms per $\alpha$-cage, namely the average number of $s$-electrons per $\alpha$-cage.  Hereafter, K-loaded K$_{12}$-A is abbreviated as K$_n$/K$_{12}$-A, and $n$ is simply called the average number of $s$-electrons per cage. The saturation loading density is $n$ $\approx\,$7.2~\cite{Nakano1999-KA}.  The $s$-electrons of the K clusters are basically confined by the negatively charged $\alpha$-cages and the positive charges of the K cations in the $\alpha$-cages, but are partly extended to adjoining $\alpha$-cages through the 8Rs. 

The simplest model of the quantum states of $s$-electrons in a K cluster in an $\alpha$-cage is given by the spherical quantum-well (SQW) potential model with an inside diameter of the $\alpha$-cage of $\approx\,$11 \AA{}, as shown in Fig.~\ref{fig:Cluster},  where the quantum states 1$s$, 1$p$, and 1$d$ appear in increasing order of energy.  Optical excitations are allowed from the 1$s$ to 1$p$ states and from the 1$p$ to 1$d$ states at  excitation energies of 1.2 and 1.5 eV, respectively, but are forbidden from the 1$s$ to 1$d$ states. $s$-electrons successively occupy these quantum states, with two occupying the 1$s$ state, the next six occupying the 1$p$ state, and so on.  According to this model, an optical excitation from the 1$s$ to 1$p$ states increases up to $n = 2$, decreases at $n > 2$, and disappears at the full occupation of 1$p$ states at $n = 8$.  A new optical excitation from the 1$p$ to 1$d$ states appears at $n > 2$.  The optical absorption and reflection spectra at room temperature have been explained by this SQW model together with the surface plasmon effect by the collective excitation of confined $s$-electrons for larger $n$ \cite{Kodaira1993-KA}. 

These aspects of alkali-metal clusters in zeolites can be compared to \textit{superatoms} \cite{Arita2004, Aoki2004, Nakano2013ICMInv} with the electronic shell structure of metal clusters \cite{Heer1987}, and their regular arrays to \textit{cluster crystals} \cite{Bogomolov1978, Nozue1992-KA, Nakano2007}.  New promising material designs have been proposed based on the regular nanostructures of porous crystals.

The first ferromagnetic properties have been observed in K$_n$/K$_{12}$-A for $2 \lessapprox n \lessapprox 7$, and were interpreted in terms of itinerant electron ferromagnetism \cite{Nozue1993-KA}. After detailed studies, an antiferromagnetic interaction between localized magnetic moments and insulating properties was suggested based on the negative values of the Weiss temperature and finite optical gaps in the infrared absorption spectra \cite{Nakano1999-KA}.  These results indicate that the system is basically an antiferromagnetic Mott insulator \cite{Nakano1999-KA}. Spontaneous magnetization is observed only for $2<n<7$ \cite{Nakano2000MCLC, Nakano2000}, and can be explained by the spin canting caused by the Dzyaloshinsky--Moriya (DM) antisymmetric exchange interaction between degenerate 1$p$ states of K clusters in adjoining $\alpha$-cages \cite{Nakano2017-APX, Nakano2004, Nakano2009muSR-KA}.  Superlattice reflections have been observed for $n>2$ in diffraction experiments \cite{Maniwa1999, Ikeda2000KA}. The average crystal structures for $n>2$ is proposed to be $F23$ \cite{Ikeda2000KA, Ikeda2004KA}.  The superlattice structure indicates that the structures of K clusters in adjoining $\alpha$-cages are not equivalent to each other, which is consistent with the necessary condition of the DM interaction that the structure lack inversion symmetry at the center of adjacent magnetic moments.  Furthermore, the degeneracy of the 1$p$ electronic states can enhance the DM interaction \cite{Tachiki1968}.  We propose an enhancement effect of the DM interaction  between adjoining $1p$ orbitals by the extension of the Rashba mechanism of spin-orbit interaction.   Theoretical calculations of electronic properties have been performed for simplified crystal structures \cite{Arita2004, Nohara2009, Nohara2011}.

A strong spin--orbit interaction (SOI) is expected at the degenerate 1$p$ states of K clusters, similarly to the Rashba mechanism \cite{Nakano2017-APX, Nakano2007}. This effect is induced by the non-orthogonality between $s$-electrons in the 1$p$ state and core electrons of the K cations located at the asymmetric potential of the K cluster. The degenerate state is usually unstable against Jahn--Teller distortion without an SOI.  The degenerate 1$p$ state of K clusters, however, can be stabilized at low temperatures due to the suppression of the Jahn--Teller effect by the strong SOI \cite{Nakano2017-APX, Nakano2007-HighMag}.

According to the SQW potential model shown in Fig.~\ref{fig:Cluster}, magnetic moments are expected to appear at odd numbers of $s$-electrons in a K cluster without Hund's coupling, such as 1, 3, 5, and 7.  Experimental results, however, indicate that the Curie constant with the occupation of $s=1/2$ spin at $\approx\,$80\% $\alpha$-cages is observed continuously for $3 \lessapprox n \leqq 7.2$ \cite{Nakano2000}.  In the present study, an optical reflection band at 0.7 eV is clearly observed for $2 < n < 6$ at low temperatures, and can be assigned to the excitation of the $\sigma$-bonding state between 1$p$-states in adjoining $\alpha$-cages. The electrical resistivity indicates the material is an insulator for all $n$.  We propose an advanced SQW model by considering $\sigma$-bonding, the orbital orthogonality of the 1$p$-hole states, and the superlattice structure. We show that the observed electronic properties can be explained by this model based on a correlated polaron system.

%%%%%%%%%%%%%%%%%%%%%%%%%%%%%%%%%%

\section{Experimental Procedures}

The zeolites studied were a crystalline powder of a few microns grain size.  Zeolite K$_{12}$-A was fully dehydrated in vacuum at 500$^\circ$C for one day.  Distilled K metal was placed in a quartz glass tube together with the dehydrated K$_{12}$-A in a glovebox filled with pure He gas containing less than 1 ppm of O$_2$ and H$_2$O.  The K metal in the quartz glass tube was adsorbed into the K$_{12}$-A at 150$^\circ$C.  The thermal annealing was continued for long enough to obtain homogeneous K-loading. The homogeneity of $n$ is within $\pm\,$0.1. The value of $n$ was estimated from the weight ratio of K-metal to K$_{12}$-A powder.

The optical diffuse reflectivity $r$ was measured by an FTIR spectrometer (Nicolet Magna 550) and a double monochromator-type UV-vis-NIR spectrometer (Varian Cary 5G). KBr powder was used as a reference white powder. Since the samples are extremely air-sensitive, the optical measurements were performed on samples sealed in quartz glass tubes.  For low temperature measurements, cryostats (Optistat DN/Bath, Oxford Instruments) were used. $r$ was transformed to the optical absorption spectrum by the Kubelka--Munk function $(1-r)^2/2r$, which gives the ratio of the absorption coefficient to the reciprocal of the powder size.  The sum of the normal reflectivity $R$ and the transmission coefficient $T_{\rm{r}}$ was obtained by the transformation $R+T_{\rm{r}} = 4r/(1+r)^2$ \cite{Kodaira1993-KA}. The normal reflectivity spectrum was obtained as $R=4r/(1+r)^2$ at the spectral region $T_{\rm{r}} \ll R$.

A SQUID magnetometer (MPMS-XL, Quantum Design) was used for magnetic measurements in the temperature range 1.8--300 K. The SQUID signal included a diamagnetic signal from the quartz glass tube as temperature-independent background, and this was subtracted from the measured magnetization. 

The electron paramagnetic resonance (EPR) or electron spin resonance (ESR) measurements were carried out using a Brucker EPR EMX spectrometer at the X-band microwave frequency ($f \approx 9.7$ GHz). The sample temperature was controlled at 3.5--300 K.

For electrical resistivity measurements, powder samples were put between two gold electrodes, and an adequate compression force of $\approx\,$1 MPa was applied during the measurements.  Because of the extreme air-sensitivity of samples, they were kept in a handmade air-proof cell and the setting procedures were completed inside the glovebox. The cell was set into a Physical Property Measurement System (PPMS, Quantum Design), and the temperature was changed between 10 and 300 K. The electrical resistivity of the cell was measured by the four-terminal method using an Agilent E4980A LCR meter at the frequency range from 20 Hz to 2 MHz under direct current (DC). The frequency dependence of the complex impedance was analyzed by a Cole--Cole plot, and the DC and 20-Hz electrical resistivity $\rho$ was obtained by multiplication by the dimensional factor (area/thickness) of the compressed powder. Due to constriction resistance \cite{Holm1967} at connections between powder particles, as well as the low filling density of the powder particles, the observed resistivity is about two orders of magnitude larger than the true value. The relative values of different samples, however, can be compared with each other within an ambiguity factor, because of the constant compression force.  The values in the present study change over several orders of magnitude. Detailed experimental procedures are explained elsewhere \cite{Nozue-Na-LSX2012}.  The upper limit of the present resistivity measurements was $\approx\,$10$^9$ $\Omega\,$cm, and obtained values of $\rho \gtrapprox$ 10$^9$ $\Omega\,$cm are unreliable. The ionic conductance of dehydrated zeolites under a low compression force is expected to be on the order of 10$^{-9}$ $\Omega^{-1}\,$cm$^{-1}$ at room temperature \cite{Kelemen1992},  and is negligible at lower temperatures in the present study. The measured value includes a small resistivity from the short circuit in the cell  ($<$ 0.1 $\Omega\,$cm), but this is negligible in the present study.

%%%%%%%%%%%%%%%%%%%%%%%%%%%%%%%%%%%

\section{Experimental Results}

\subsection{\label{sec:Optical}Optical Properties}

\begin{figure}[ht]
\resizebox*{6.8cm}{!}{\includegraphics{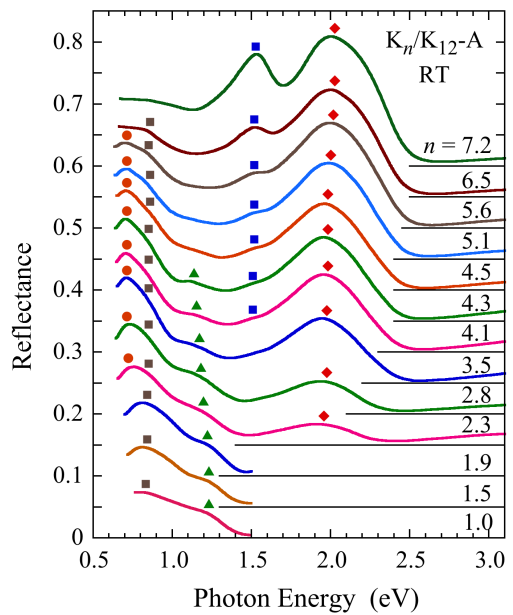}}
\caption{\label{fig:Ref-RT}(Color online) Reflection spectra of K$_n$/K$_{12}$-A at room temperature (RT). The value of $n$ is indicated for each spectrum. The colored shapes indicate discernable special features, explained in the main text.}
\end{figure}

Optical resonant absorption and reflection spectra of the allowed transitions between quantum states provide important information on electronic states, including nonmagnetic states.  The reflection spectra of K$_n$/K$_{12}$-A have been measured at room temperature (RT) in Ref.~[\onlinecite{Kodaira1993-KA}]. In the present study, we made equivalent measurements on our samples with sufficient homogenization of the K-loading, as shown in Fig.~\ref{fig:Ref-RT}. Basically, the same reflection spectra were obtained at RT.  Different structures, marked in the figure, appear depending on $n$.  In Ref.~[\onlinecite{Kodaira1993-KA}], the structures at 0.7, 0.83, and $\approx$1.2 eV were assigned to the 1$s$--1$p$ excitations of K clusters in $\alpha$-cages,  although the 0.7 and 0.83 eV bands are far from the expected energy 1.2 eV in Fig.~\ref{fig:Cluster}. The assignments of 0.7 and 0.83 eV bands are changed in the present paper, as explained later. The peaks at 1.5 and 2 eV were assigned to the 1$p$--1$d$ excitation and the surface plasmon band due to collective excitation of $s$-electrons in K clusters, respectively.  The surface plasmon resonance band dominates the spectrum at large $n$, because the oscillator strength of the collective excitation increases in K cluster with many $s$-electrons in a deep potential of an $\alpha$-cage.  Individual excitations between quantum states are relatively suppressed by the collective excitation because of the sum rule of the oscillator strength.  If K clusters were formed in $\beta$-cages, a resonance band would be expected at $\approx\,$3 eV.  However, no structure is seen around 3 eV in Fig.~\ref{fig:Ref-RT}, indicating that no clusters are formed in the $\beta$-cages.

\begin{figure}[ht]
\resizebox*{6.8cm}{!}{\includegraphics{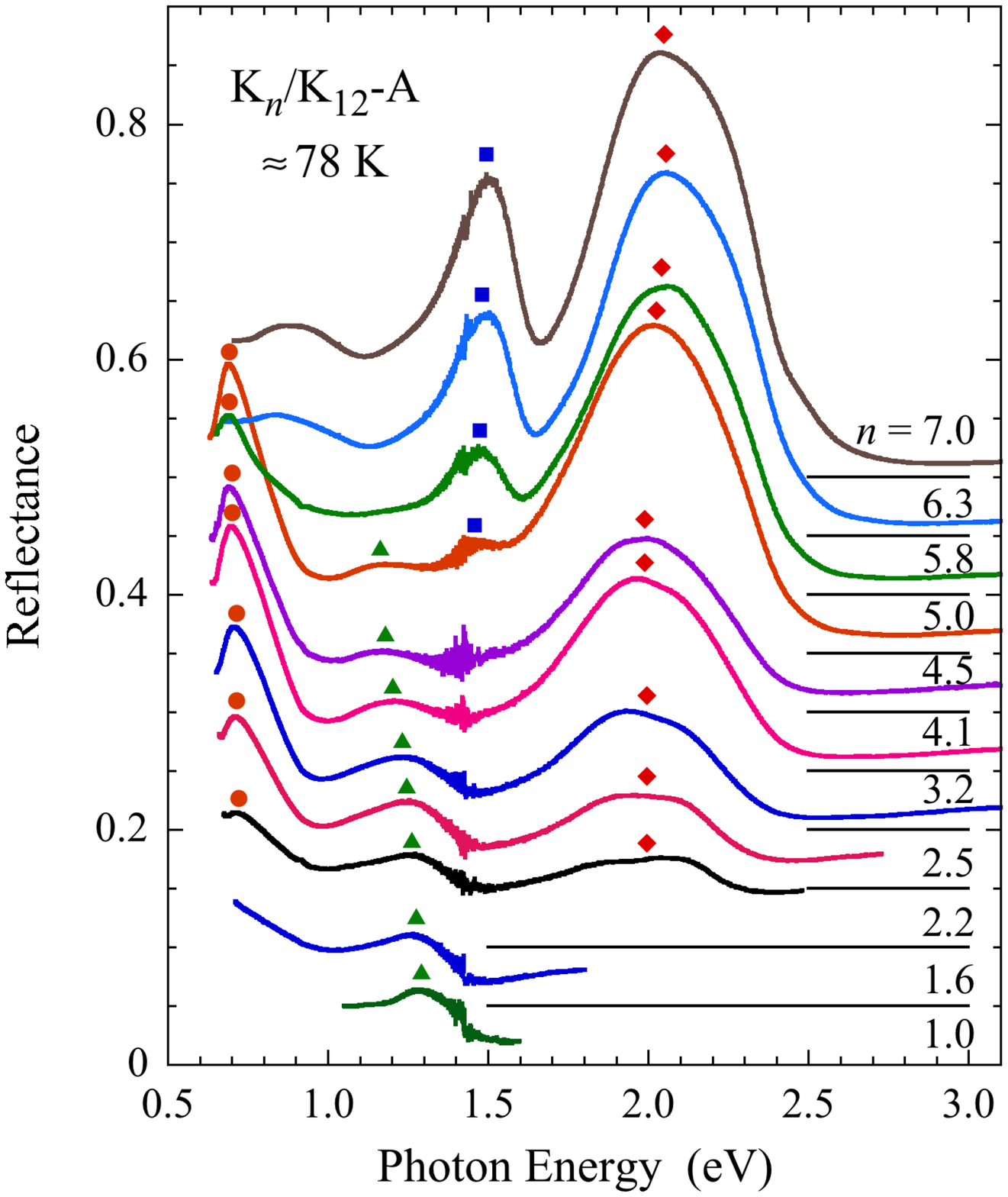}}
\caption{\label{fig:Ref-LT}(Color online) Reflection spectra of K$_n$/K$_{12}$-A at $\approx\,$78 K. The value of $n$ is indicated for each spectrum. The colored shapes indicate discernable special features, explained in the main text.}
\end{figure}

\begin{figure}[ht]
\resizebox*{7.0cm}{!}{\includegraphics{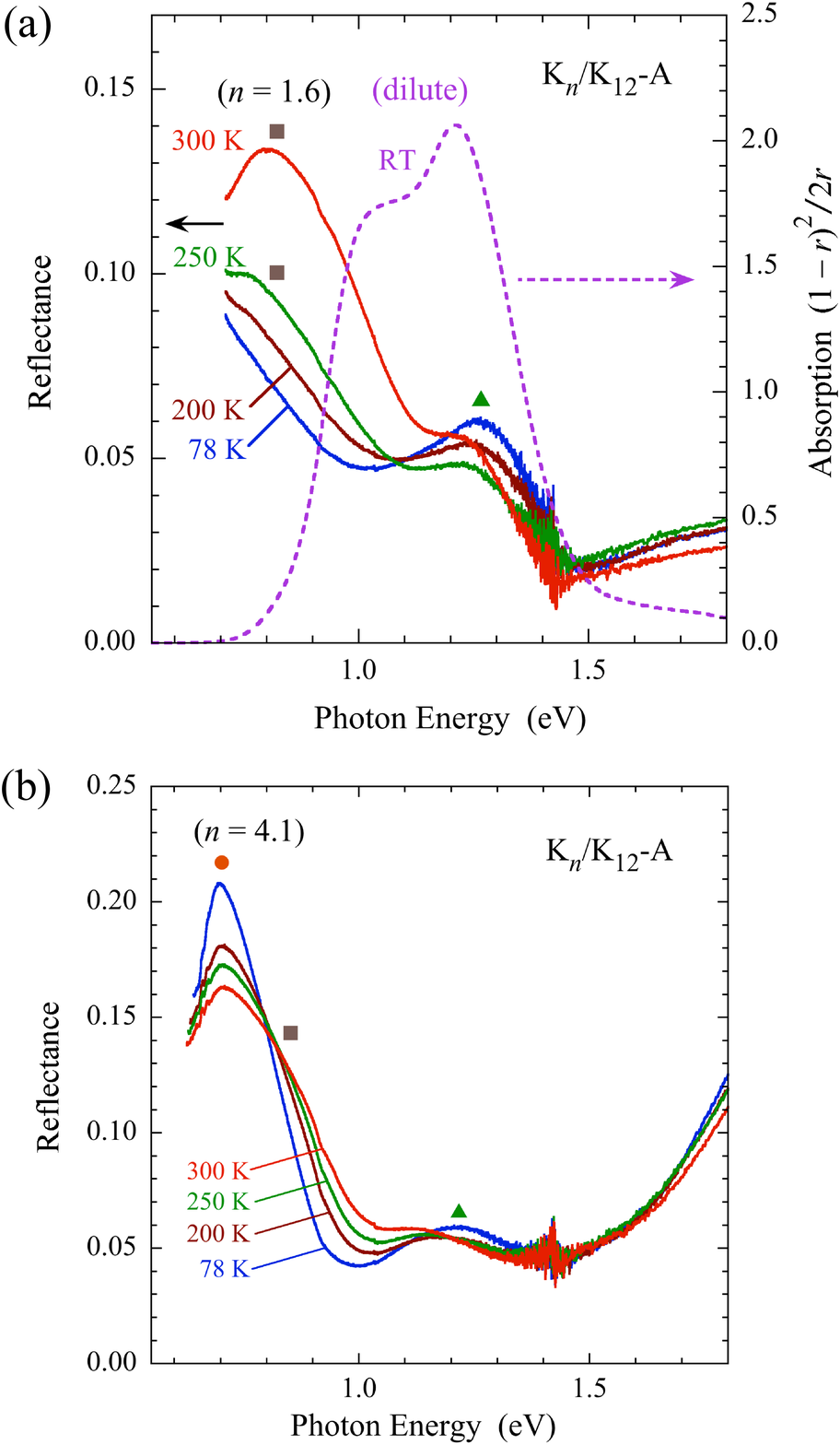}}
\caption{\label{fig:Ref1.6}(Color online) (a) Temperature dependence of reflection spectrum of K$_n$/K$_{12}$-A at $n=1.6$. An absorption spectrum of dilute K-loaded K$_{12}$-A at RT is included as a dotted curve for comparison. (b) Temperature dependence of reflection spectrum of K$_n$/K$_{12}$-A at $n=4.1$. The colored shapes indicate discernable special features, explained in the main text.}
\end{figure}

In the present study, reflection spectra at $\approx\,$78 K were measured, shown in Fig.~\ref{fig:Ref-LT}.   An increase in reflectivity at $\approx\,$78 K is observed overall, in comparison to those at room temperature at the similar $n$ in Fig.~\ref{fig:Ref-RT}.  This is because the damping of electronic states by thermal vibrations of atoms decreases at low temperatures, and the reflectivity increases.  A clear difference is seen at structures below 1 eV in Fig.~\ref{fig:Ref-LT}.  The reflection peak at 0.7 eV clearly appears only for $2 < n < 6$.  The previous assignment of the 0.7 eV band as the 1$s$--1$p$ transition in Ref.~[\onlinecite{Kodaira1993-KA}] is not applicable, because of its absence for $n<2$.  A further difference at $\approx\,$78 K is an absence of a reflection band at $\approx\,$0.83 eV.  According to the $n$ and temperature dependences, the origin of the 0.7 eV band is different from that of 0.83 eV band. In Section~\ref{sec:Electronic1p}, the origin of the 0.7 eV band is assigned to the excitation from the $\sigma$-bonding between 1$p$-states in adjoining $\alpha$-cages to its anti-bonding.  

The temperature dependence of the reflection spectra was measured at $n=1.6$, shown in Fig.~\ref{fig:Ref1.6}(a). An absorption spectrum of dilute K-loaded K$_{12}$-A at RT is indicated for comparison.    The reflection band at $\approx\,$0.83 eV becomes weak with a decrease in temperature. This band is assigned to the excitation of $s$-electrons in thermally metastable states, and is not important in the discussion at low temperatures, although the thermally metastable states are interesting nature.  The reflection band at $\approx\,$1.2 eV at RT coincides well with the optical absorption band observed in dilute K-loaded K$_{12}$-A. The energy of this band is very close to the 1$s$--1$p$ energy difference of 1.2 eV in the SQW potential model in Fig.~\ref{fig:Cluster}. The temperature dependence of the reflection spectra at $n=4.1$ is shown in Fig.~\ref{fig:Ref1.6}(b).  The reflection peak at 0.7 eV clearly appears with the similar oscillator strength at all temperatures, but the reflection band at $\approx\,$0.83 eV becomes weak with a decrease in temperature.  Because the ferromagnetic properties are observed at low temperatures, we refer to the reflection spectra at $\approx\,$78 K in Fig.~\ref{fig:Ref-LT}, instead of those at room temperature in Fig.~\ref{fig:Ref-RT}.

%%%%%%%%%%%%%%%%%%%%%%%%%%%%%%%%%%%
\subsection{\label{sec:Magnetic}Magnetic Properties}

\begin{figure}[ht]
\resizebox*{6.3cm}{!}{\includegraphics{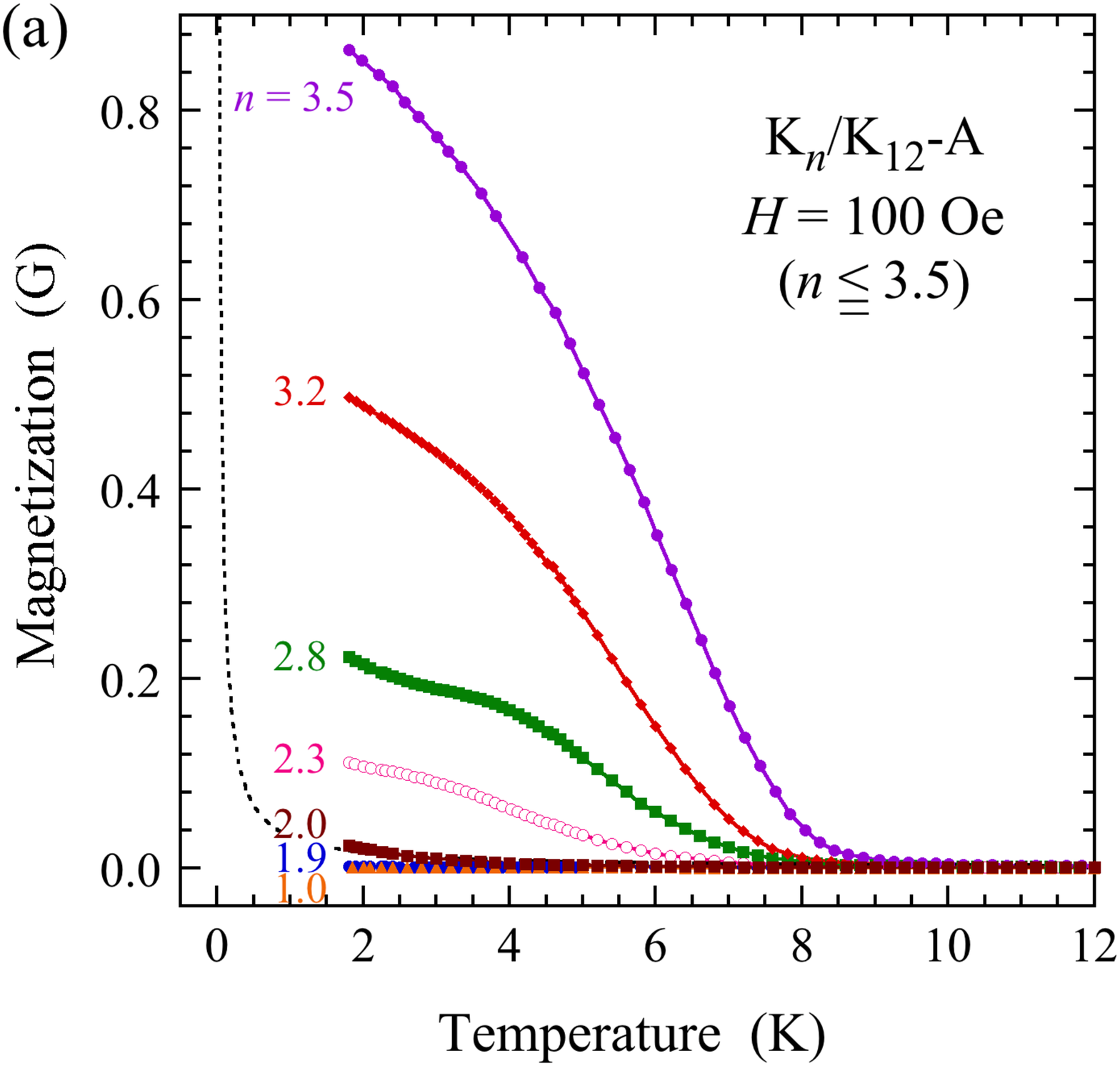}}
\resizebox*{6.3cm}{!}{\includegraphics{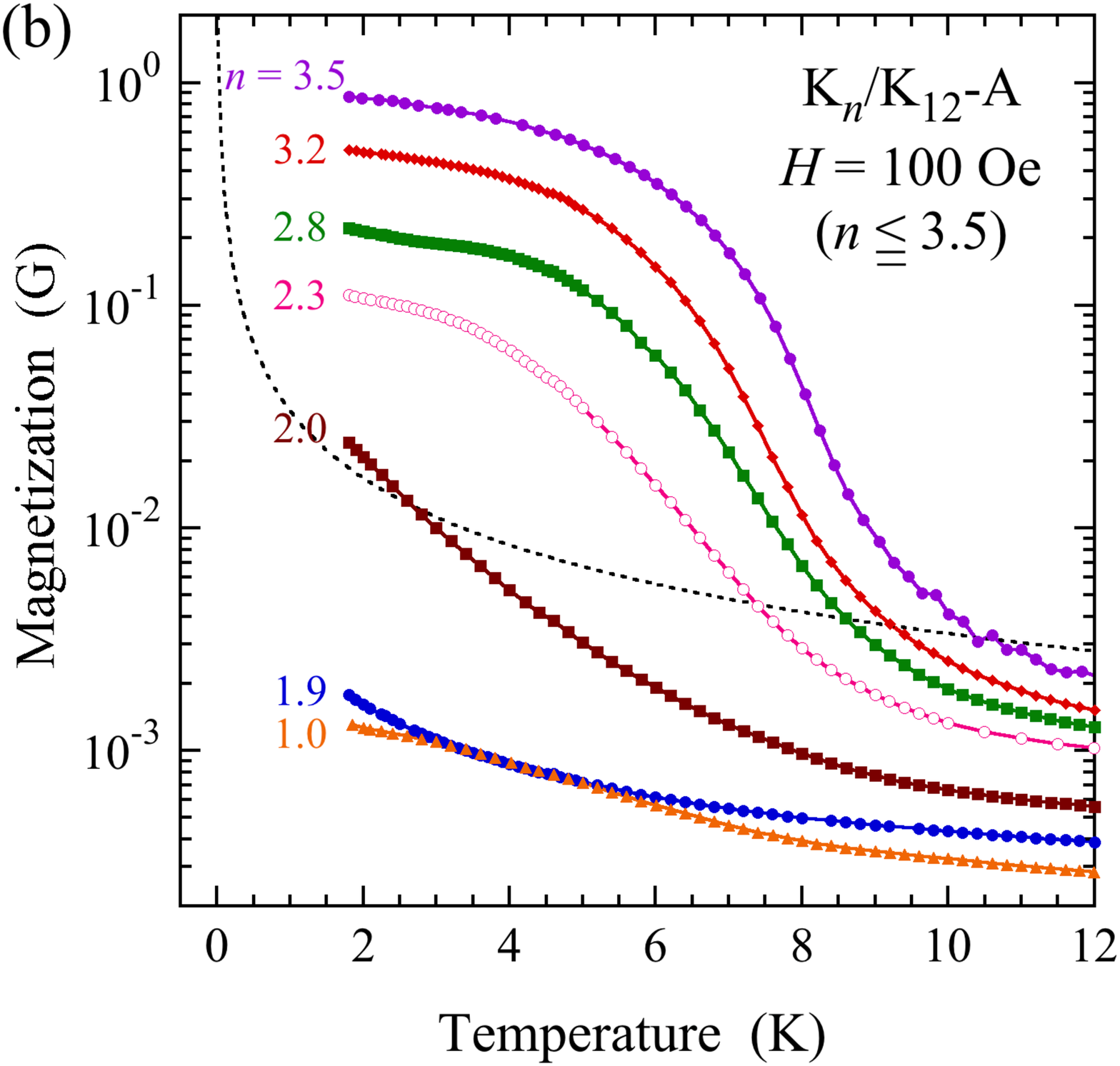}}
\caption{\label{fig:MT-1}(Color online) Temperature dependence of magnetization in K$_n$/K$_{12}$-A under a magnetic field of $H=100$ Oe for $n \leqq 3.5$ on (a) linear and (b) logarithmic scales.   The dotted curves indicate the paramagnetic magnetization at $H=100$ Oe for the Curie law of spin $s=1/2$ magnetic moments with 100\% population in the $\alpha$-cages.}
\end{figure}

\begin{figure}[ht]
\resizebox*{6.3cm}{!}{\includegraphics{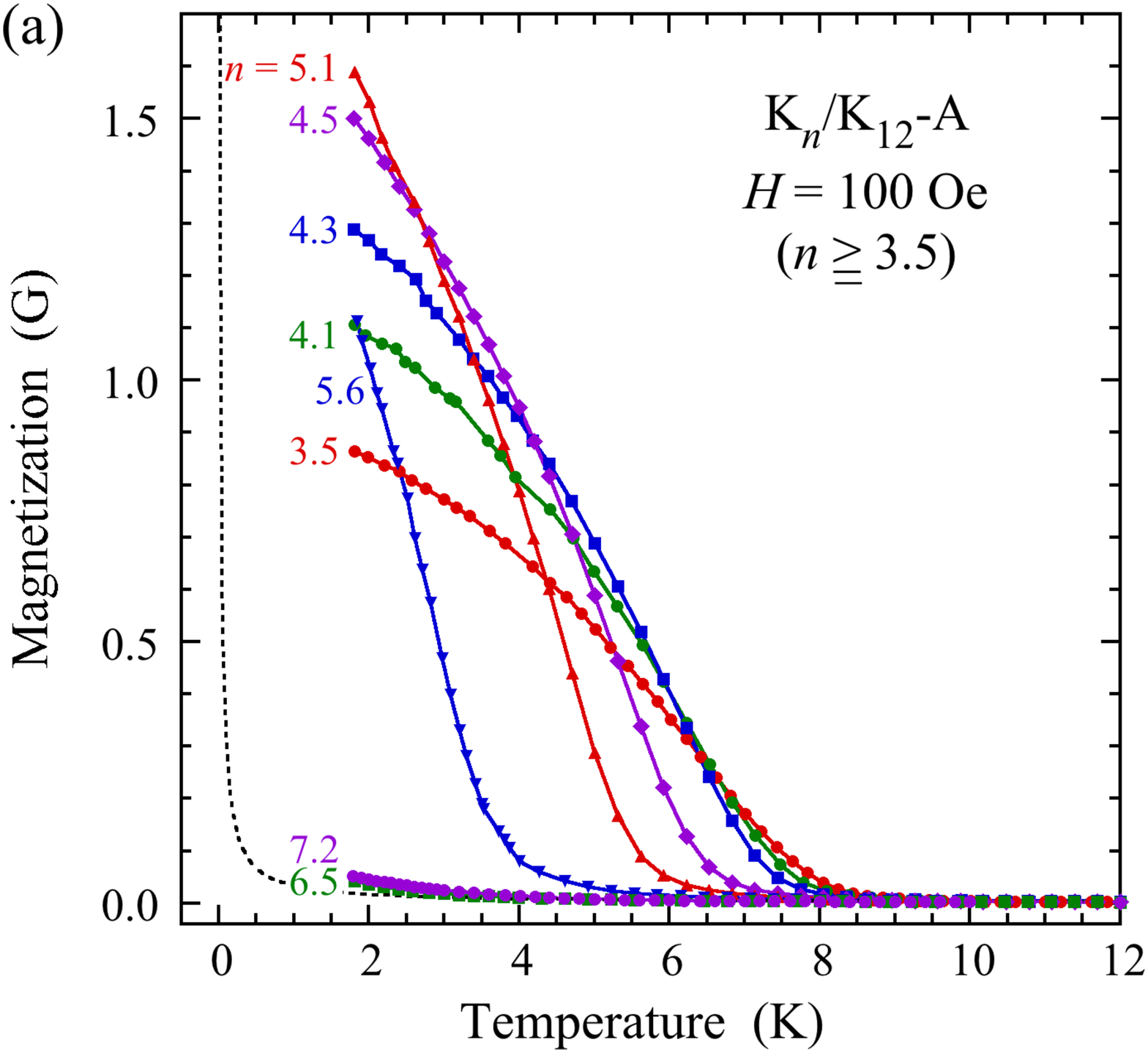}}
\resizebox*{6.3cm}{!}{\includegraphics{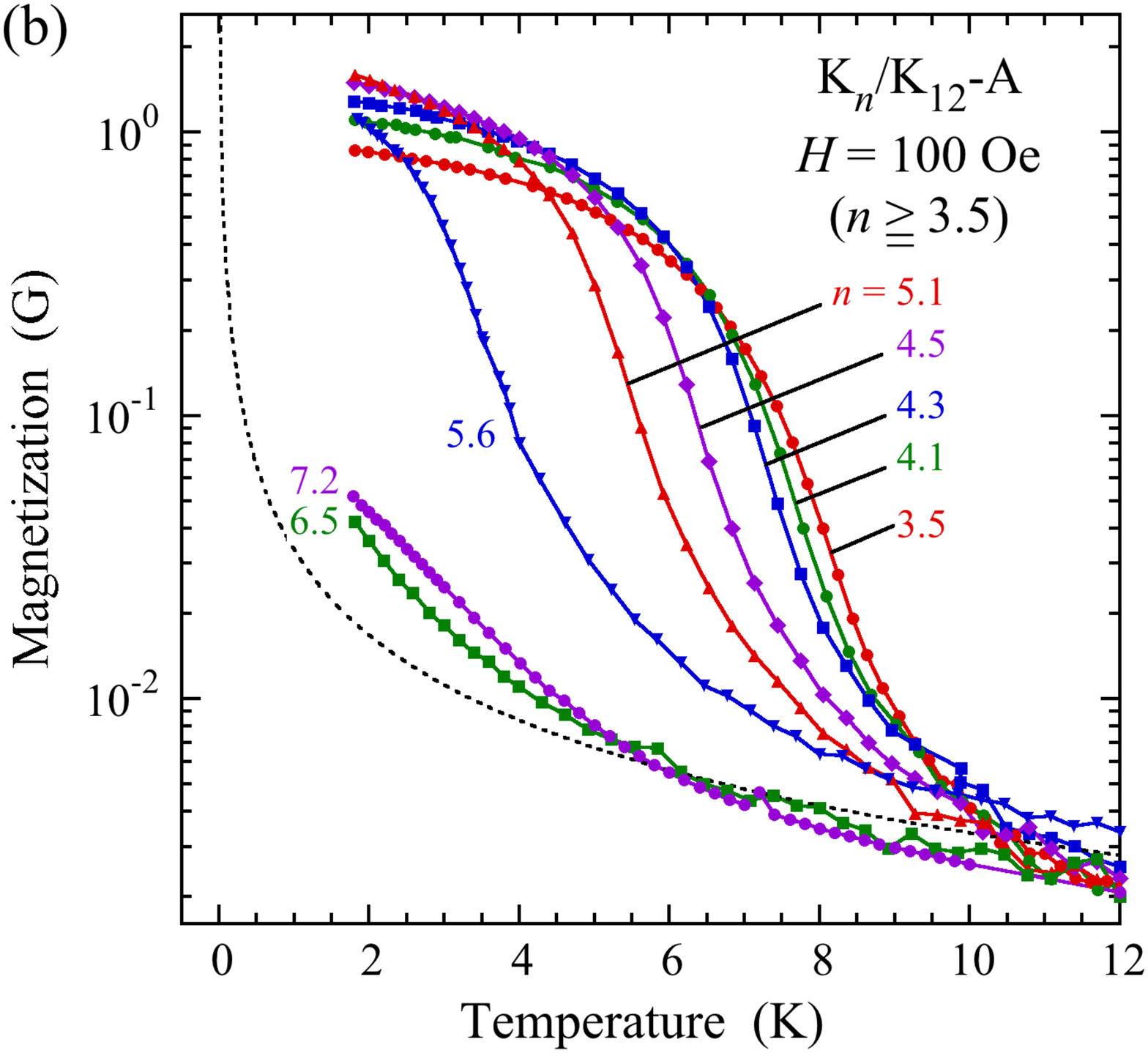}}
\caption{\label{fig:MT-2}(Color online) Temperature dependence of magnetization in K$_n$/K$_{12}$-A under a magnetic field of $H=100$ Oe for $n \geqq 3.5$ in (a) linear and (b) logarithmic scales.   The dotted curves indicate the paramagnetic magnetization at $H=100$ Oe for the Curie law of spin $s=1/2$ magnetic moments with 100\% population in the $\alpha$-cages.}
\end{figure}

The first observation of ferromagnetic properties in K$_n$/K$_{12}$-A was provided by AC magnetic susceptibility using a Hartshorn inductance bridge \cite{Nozue1992-KA, Nozue1993-KA}.  In the present study, the static magnetization was measured by a SQUID magnetometer. The temperature dependence of the magnetization under a DC magnetic field of $H=100$ Oe is shown for $n \leqq 3.5$ in Figs.~\ref{fig:MT-1}(a) and \ref{fig:MT-1}(b) on linear and logarithmic scales, respectively, and for $n \geqq 3.5$ in Figs.~\ref{fig:MT-2}(a) and \ref{fig:MT-2}(b).  The dotted curve in each figure indicates the expected paramagnetic magnetization at $H=100$ Oe from the Curie law of spin $s=1/2$ magnetic moments with 100\% population in $\alpha$-cages. The magnetization increases monotonically with decreasing temperature, which is slightly different from the AC susceptibility in Ref.~[\onlinecite{Nozue1993-KA}].  The temperature dependences for $2 < n \leqq 3.5$ in Fig.~\ref{fig:MT-1}(b) are similar to each other, indicating that the ferromagnetic volume with similar Curie temperature (slightly increasing with $n$) seems to increase with $n$.   The temperature dependences for $n \geqq 3.5$  in Fig.~\ref{fig:MT-2}(b) shift to lower temperatures with $n$, indicating that the Curie temperature decreases with $n$.  The Curie temperatures at $n=6.5$ and 7.2 are lower than the measured temperature 2 K or are paramagnetic down to 0 K.  The magnetization at $n=7.2$ has a very small ferromagnetic component at low temperatures, but almost follows the Curie law ($T_{\rm{C}}=0$) of isolated spins $s=1/2$ with $\approx\,$80\% population in $\alpha$-cages, as mentioned in Ref.~[\onlinecite{Nakano1999-KA}].  The temperature dependences of the magnetization for $2 < n \leqq 3.5$  in Fig.~\ref{fig:MT-1} are quite different from those for $n \geqq 3.5$  in Fig.~\ref{fig:MT-2}. 

\begin{figure}[ht]
\resizebox*{6.0cm}{!}{\includegraphics{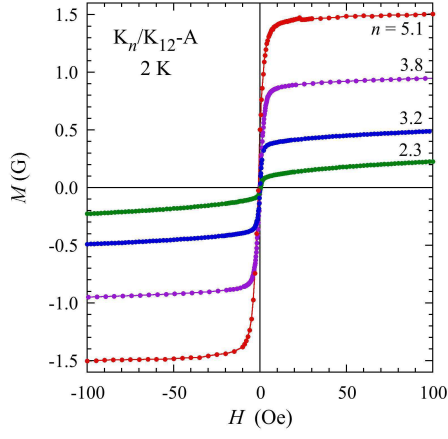}}
\caption{\label{fig:MH-fine}(Color online) Magnetization curve in K$_n$/K$_{12}$-A at low magnetic fields at 2 K.   The value of $n$ is indicated for each curve.  A hysteresis was less than the detection limit.}
\end{figure}

\begin{figure}[ht]
\resizebox*{7.0cm}{!}{\includegraphics{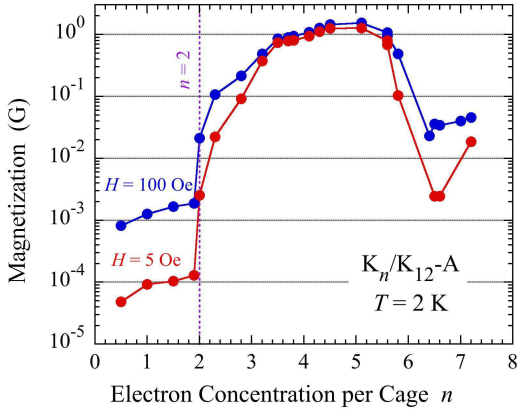}}
\caption{\label{fig:M5Oe100Oe}(Color online) $n$-dependences of magnetization at $H = 5$ and 100 Oe at 2 K in K$_n$/K$_{12}$-A.  }
\end{figure}

\begin{figure}[ht]
\resizebox*{7.5cm}{!}{\includegraphics{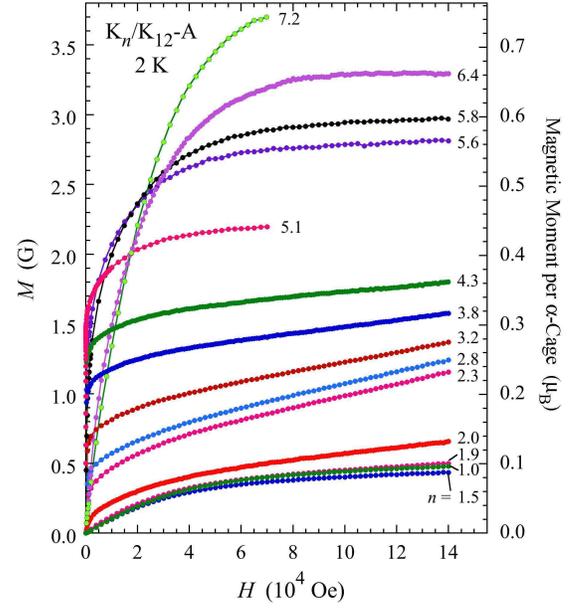}}
\caption{\label{fig:MH-14T}(Color online) Magnetization curve in K$_n$/K$_{12}$-A up to high magnetic fields at 2 K.  The value of $n$ is indicated for each curve.}
\end{figure}

The magnetization curve of ferromagnetic samples  are shown for K$_n$/K$_{12}$-A at low magnetic fields at 2 K in Fig.~\ref{fig:MH-fine}.   A steep increase in magnetization is seen around zero field.  A hysteresis was less than the detection limit, and the ferromagnetic property is very soft~\cite{Nozue1993-KA, Nakano2009muSR-KA}.  Such a soft magnetization is generally observed in the spin-cant antiferromagnet, if the external field is applied along the easy plain~\cite{Nakano2009muSR-KA, Pincus1960}.  Present sample  particles, however, are randomly oriented. Magnetic domain walls move easily by the any direction of external magnetic field.  Easy plains scarcely exist.   The $n$-dependences of magnetization at $H = 5$ and 100 Oe at 2 K in K$_n$/K$_{12}$-A are shown in Fig.~\ref{fig:M5Oe100Oe}.  A steep increase in several orders of magnitude is seen at just above $n = 2$.    

The magnetization curve in K$_n$/K$_{12}$-A up to high magnetic fields at 2 K are shown in Fig.~\ref{fig:MH-14T}.  A steep increase near zero fields are seen in ferromagnetic samples. A gradual increase is seen at medium fields. A constant increase is seen in ferromagnetic samples at high magnetic field.   At much higher magnetic fields up to $52 \times 10^4$ Oe, an anomalous magnetization has been observed because of the spin multiplet term of 1$p$ state \cite{Nakano2017-APX, Nakano2007-HighMag, Nakano2007}. 

\begin{figure}[ht]
\resizebox*{7.0cm}{!}{\includegraphics{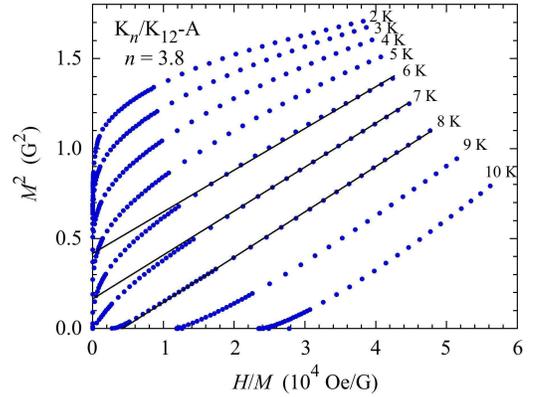}}
\caption{\label{fig:Arrott_n38}(Color online) Arrott plot analysis at $n = 3.8$ in K$_n$/K$_{12}$-A. The temperature is indicated for each curve. }
\end{figure}

An Arrott plot analysis is shown for K$_n$/K$_{12}$-A at $n = 3.8$ in Fig.~\ref{fig:Arrott_n38}.  The existence of spontaneous magnetization can be confirmed below 7 K.  The Curie temperature is estimated from the temperature at which the strait line intercepts the origin.  An existence of macroscopic magnetization has been confirmed by the muSR experiment \cite{Nakano2009muSR-KA}.

\begin{figure}[ht]
\resizebox*{6.5cm}{!}{\includegraphics{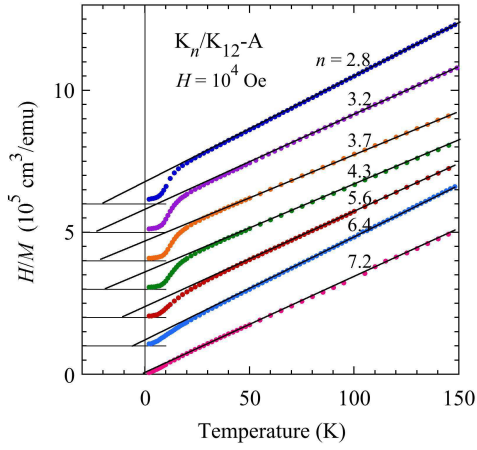}}
\caption{\label{fig:CW-n}(Color online) Curie-Weiss behavior of the reciprocal of magnetic susceptibility in K$_n$/K$_{12}$-A. The value of $n$ is indicated for each curve. }
\end{figure}

The temperature dependence of the reciprocal of magnetic susceptibility in K$_n$/K$_{12}$-A is plotted in Fig.~\ref{fig:CW-n}.  A Curie-Weiss law is seen at higher temperatures. At low temperatures, a deviation from the Curie-Weiss law is seen, because of ferromagnetic properties.  The Weiss temperature in ferromagnetic samples is negative, indicating that the antiferromagnetic interaction between magnetic moments dominates the system, although ferromagnetic properties are observed.  A candidate of such magnetism is ferrimagnetism or spin-cant of antiferromagnetism.  According to the magnetization up to high magnetic fields, the origin of magnetism has been assigned to the spin-cant of antiferromagnetism~\cite{Nakano2004}.

\begin{figure}[ht]
\resizebox*{7cm}{!}{\includegraphics{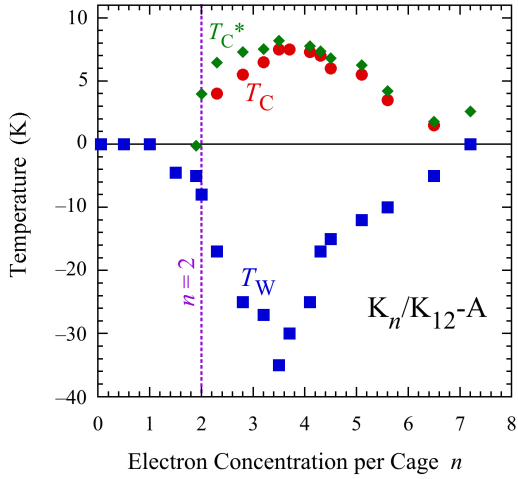}}
\caption{\label{fig:Curie}(Color online) $n$-dependence of the Curie temperature $T_{\rm{C}}$, asymptotic Curie temperature $T_{\rm{C}}$* and Weiss temperature $T_{\rm{W}}$ in K$_n$/K$_{12}$-A. }
\end{figure}

We estimate the asymptotic Curie temperature $T_{\rm{C}}$* from the temperature dependence of the magnetization, shown in Fig.~\ref{fig:Curie} as a function of $n$. The Curie temperature $T_{\rm{C}}$ estimated from an Arrott-plot analysis and the Weiss temperature $T_{\rm{W}}$ estimated from the Curie-Weiss law \cite{Nakano2001-RbA-KA} are also shown.  $T_{\rm{C}}$ suddenly appears for $n > 2$, increases with $n$, has a peak of $\approx\,$8 K at $n \approx 3.6$, and decreases to 0 K at $n = 7.2$. $T_{\rm{W}}$ suddenly decreases for $n > 2$ to a negative value, decreases with $n$, has a minimum of $\approx\,$$-35$ K at $n \approx 3.6$, and increases to 0 K at $n = 7.2$.   If the ferromagnetic properties are inhomogeneous, $T_{\rm{C}}$ reflects the main part of the spontaneous magnetization, while $T_{\rm{C}}$* is affected by the minority ferromagnetic part with higher Curie temperature.   In Fig.~\ref{fig:Curie}, $T_{\rm{C}}$* is much higher than $T_{\rm{C}}$ for $2<n<3.5$, but similar to $T_{\rm{C}}$ for $3.5<n<7$.  A large inhomogeneity in the ferromagnetic properties is expected for $2<n<3.5$.  At $n=7.2$, $T_{\rm{C}}$* is much higher than $T_{\rm{C}}=0$ K in Fig.~\ref{fig:Curie}, because of the very small ferromagnetic component shown in Fig.~\ref{fig:MT-2}(b).

\begin{figure}[ht]
\resizebox*{7.0cm}{!}{\includegraphics{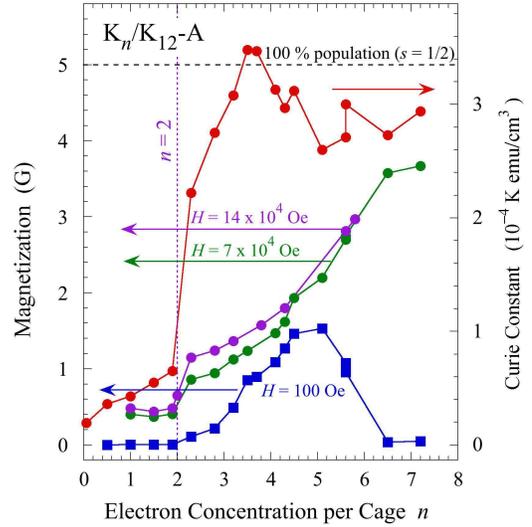}}
\caption{\label{fig:CurieConst}(Color online) $n$-dependence of magnetization at 2 K in K$_n$/K$_{12}$-A for $H=$ 100, $7 \times 10^4$ and $14 \times 10^4$ Oe, and $n$-dependence of the Curie constant.  The horizontal dotted line is the expected value of the saturation magnetization or the Curie constant for 100\% population of spin $s=1/2$ magnetic moments in the $\alpha$-cages. }
\end{figure}

\begin{figure}[ht]
\resizebox*{6.2cm}{!}{\includegraphics{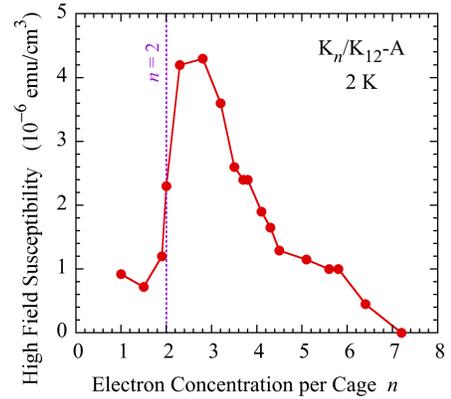}}
\caption{\label{fig:hf-chi}(Color online) $n$-dependence of high-field magnetic susceptibility estimated from the slope of the magnetization curve for $H > 6 \times 10^4$ Oe at 2 K in K$_n$/K$_{12}$-A. }
\end{figure}

The $n$-dependence of the magnetization at 2 K is plotted for $H=14 \times 10^4$ Oe in Fig.~\ref{fig:CurieConst}, together with those at $H=$ 100 and $7 \times 10^4$ Oe and the $n$-dependence of the Curie constant \cite{Nakano2000}.  A horizontal dotted line shows the Curie constant ($3.35 \times 10^{-4}$ K\,emu/cm$^3$) of spin $s=1/2$ magnetic moments with 100\% population in the $\alpha$-cages and the corresponding saturation magnetization (5.0 G). The magnetization ($\gtrapprox$ 0.1 G) at $H =100$ Oe for $2 < n < 6$ originates from the spontaneous magnetization.  The very small magnetization for $n>6$ is mainly caused by the paramagnetism at 2 K.  The magnetization at $H =100$ Oe  suddenly increases for $n>2$, has a step at $n \approx 3.5$, and has a peak at $n \approx 5$. The Curie constant suddenly increases for $n>2$, has a peak at $n \approx 3.6$, and keeps the value of $\approx80$\% population of spin $s=1/2$ magnetic moments in the $\alpha$-cages up to $n = 7.2$. 

The magnetization curve up to $H = 7 \times 10^4$ Oe at 2 K has four parts: (1) the spontaneous magnetization of ordered moments, (2) an increase in the spin-cant angle of ordered moments by the applied magnetic field, (3) suppression of the thermal fluctuation of ordered moments by the applied magnetic field, and (4) magnetization of isolated paramagnetic moments partly distributed in the sample.  At high magnetic fields of $H = 7 \times 10^4$ and $14 \times 10^4$ Oe, the observed magnetizations for $2 < n < 6$ in Fig.~\ref{fig:CurieConst} are much smaller than the saturation value estimated from the Curie constant, indicating that the magnetization is suppressed by an antiferromagnetic interaction between magnetic moments. Magnetizations for $n > 6$, however, approach the saturation value estimated from the Curie constant, because 2 K is a paramagnetic temperature.  This situation is clearly seen in the high-field magnetic susceptibility estimated from the slope of the magnetization curve for $H > 6 \times 10^4$ Oe at 2 K shown in Fig.~\ref{fig:hf-chi}.  The high-field susceptibility suddenly increases for $n>2$, has a peak around $n \approx 2.8$, and decreases gradually with $n$.  The high-field magnetic susceptibility is explained by the increase in the cant-angle under the applied magnetic field.  The decrease in the high-field magnetic susceptibility with the increase in $n$ is explained by the increase in the DM interaction \cite{Nakano2004}.

At $n=7.2$, a Curie law ($T_{\rm{C}} = T_{\rm{W}} = 0$) is observed, although most of the $\alpha$-cages ($\approx\,$80\%) are occupied by $s = 1/2$ moments. This result indicates that adjacent clusters have no magnetic interaction \cite{Nakano1999-KA}. This result can be explained by mutually orthogonalized nondegenerate 1$p$-holes \cite{Nakano1999-KA, Nakano2002ESR}, and is discussed in Section~\ref{sec:Electronic1p}. 

At much higher magnetic fields up to $H=5.2 \times 10^5$ Oe, extraordinary magnetization has been observed at $n=4.5$ \cite{Nakano2017-APX,Nakano2007-HighMag}. The magnetic moment  per cage at 1.3 K amounts to 1.4$\mu_{\rm{B}}$ which exceeds the value 0.96$\mu_{\rm{B}}$ estimated from the Curie constant. The extraordinary magnetic moment can be explained by a crossover between degenerate 1$p$(E$_u$) states with angular momenta of $J_z = \pm 3/2$ and $\pm 1/2$ at high magnetic fields. The spin--orbit splitting energy is estimated to be $\approx$1.5 meV \cite{Nakano2007-HighMag},  which corresponds well with a roughly estimated value \cite{Nakano2007}.  On the other hand, the magnetization at $n=7.2$ saturates at high magnetic fields, and does not exhibit any extraordinary magnetization \cite{Nakano2017-APX}. 

\begin{figure}[t] %aaaaa
%\begin{center}
\resizebox*{4.0cm}{!}{\includegraphics{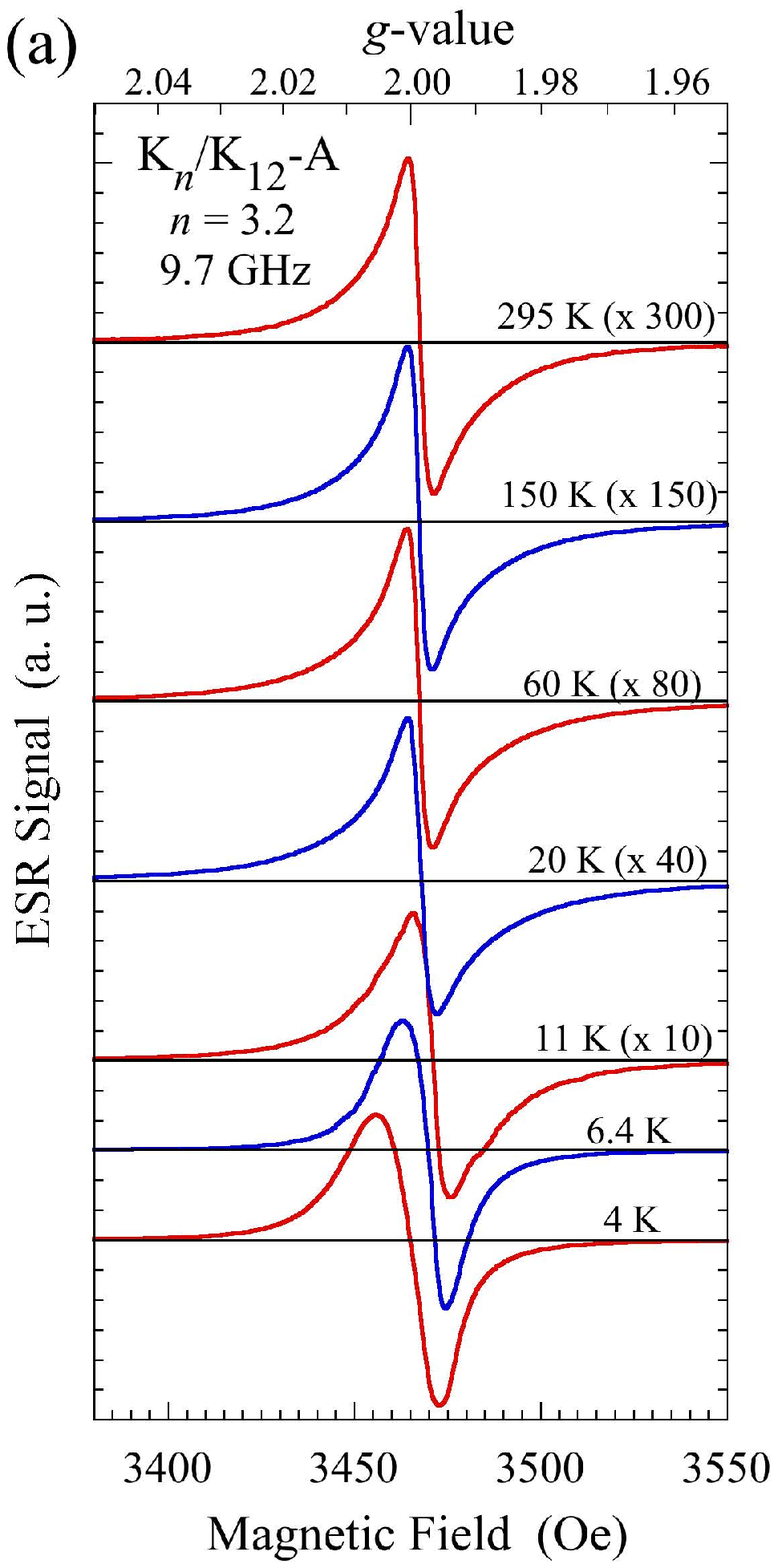}}
\resizebox*{4.22cm}{!}{\includegraphics{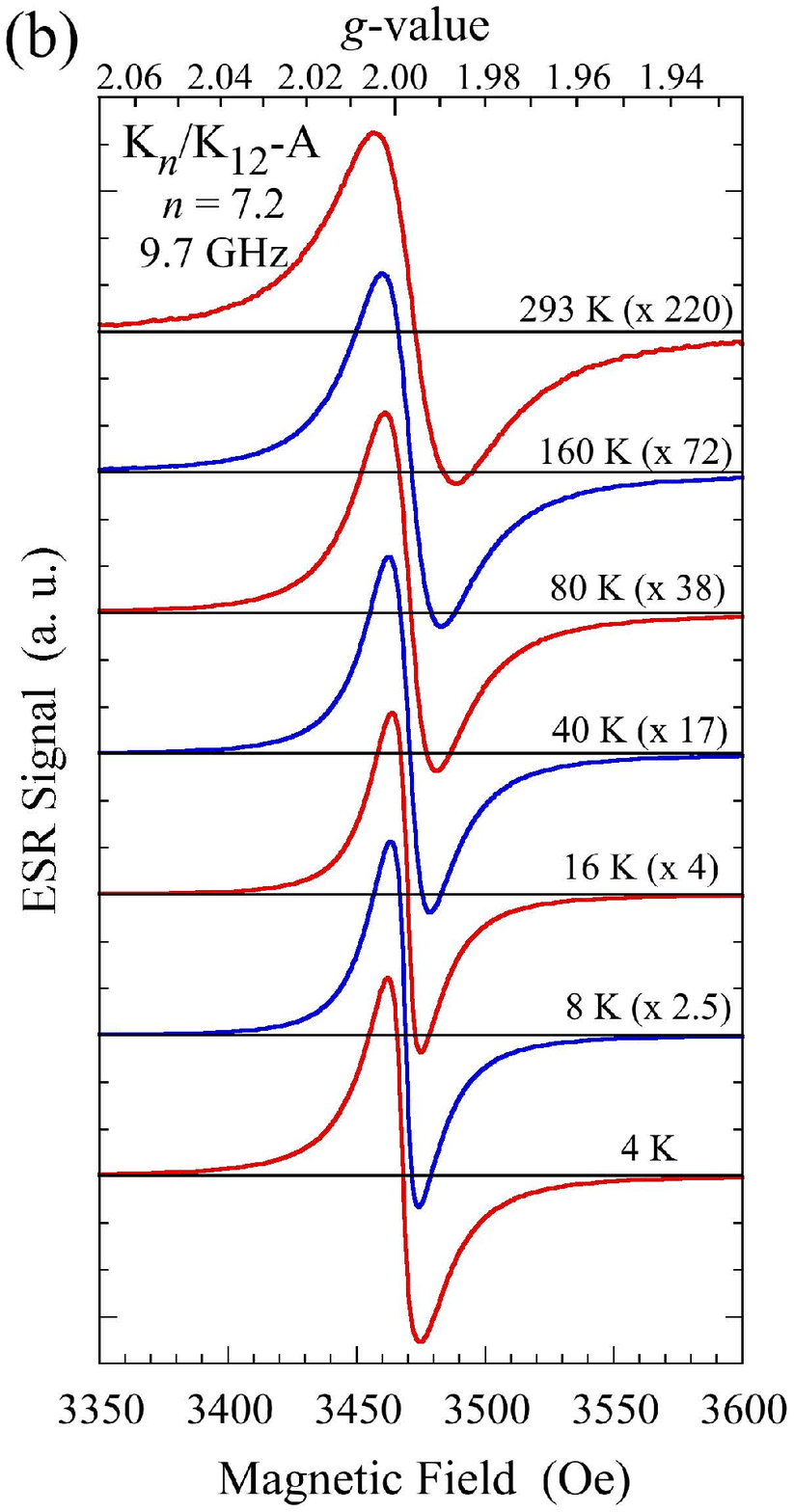}}
\caption{(Color online) Temperature dependences of ESR spectrum in K$_n$/K$_{12}$-A at (a) $n = 3.2$ and (b) $n = 7.2$. Temperature is indicated for each curve.  The magnetic field is normalized at 9.7 GHz to have a common $g$-value which is indicated at the top of figure. }
\label{fig:ESR-spect}
%\end{center}
\end{figure}

\begin{figure}[ht]
\resizebox*{7cm}{!}{\includegraphics{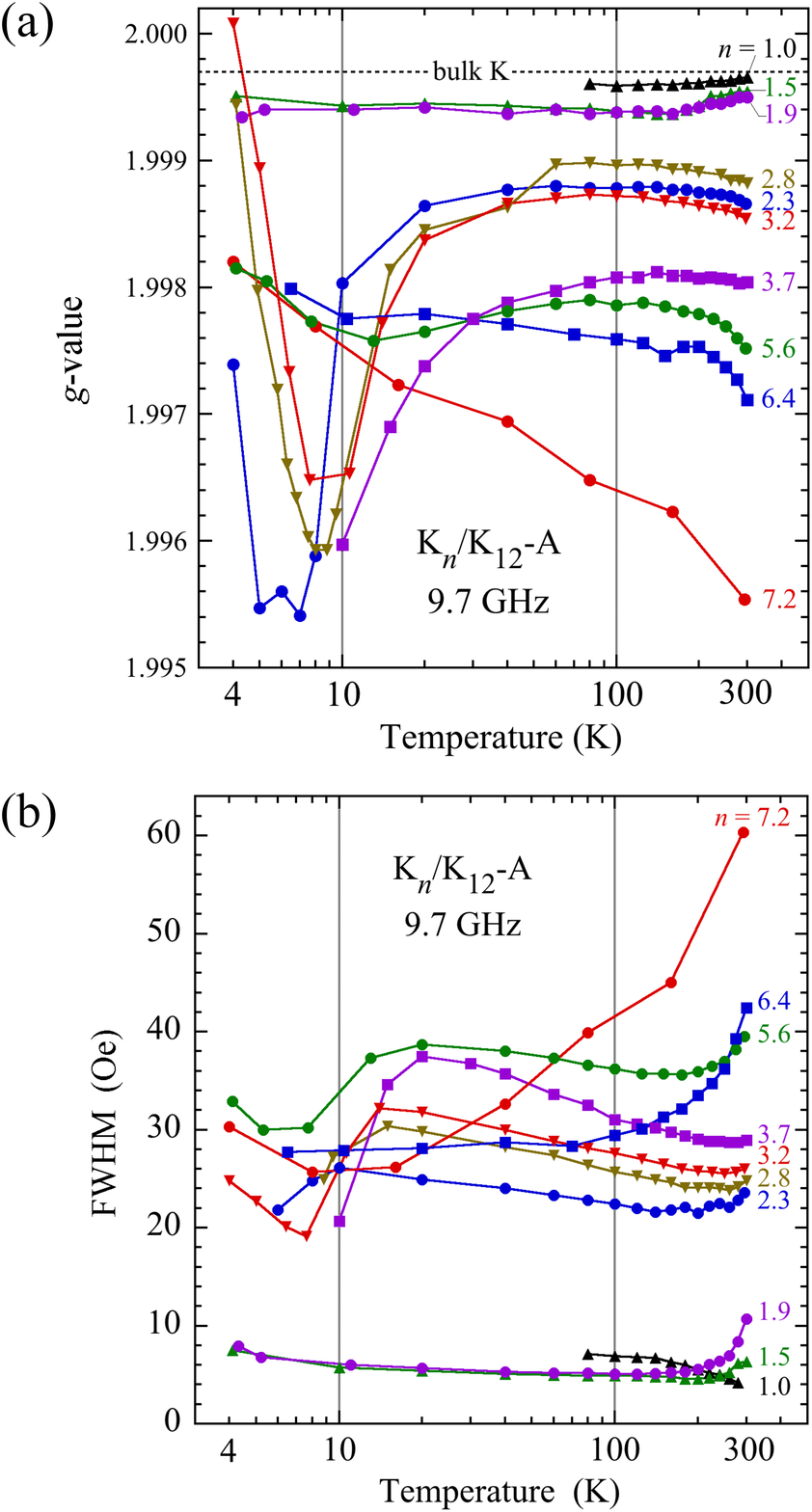}}
\caption{\label{fig:(K_KA)g-T}(Color online) (a) Temperature dependence of $g$-value in EPR (or ESR) spectrum at various values of $n$ in K$_n$/K$_{12}$-A.  The horizontal dotted line shows the $g$-value of bulk K metal.  (b) Temperature dependence of FWHM in EPR (or ESR) spectrum at various values of $n$ in K$_n$/K$_{12}$-A.}
\end{figure}

\begin{figure}[ht]
\resizebox*{6cm}{!}{\includegraphics{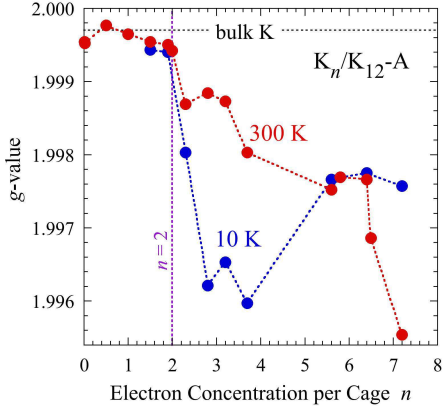}}
\caption{\label{fig:g-10K-n}(Color online)  $n$-dependences of $g$-value in EPR spectrum at 10 and 300 K in K$_n$/K$_{12}$-A.  The horizontal dotted line shows the $g$-value of bulk K metal.  }
\end{figure}

%%%%%%%%%%%%%%%%%%%%%%%%%%%%%%%%%%%
\subsection{\label{sec:ESR}ESR Properties}

The EPR spectrum of an isolated spin has a hyperfine structure due to the interaction with nuclear spins.   The hyperfine structure can also exhibit motional narrowing and/or exchange narrowing in solids.  The EPR in isotropic solids has a finite spectral width due to relaxation by the spin--phonon interaction, and also shows a decrease in $g$-value due to the contribution of the orbital angular momentum to the magnetic moment.  Both the increase in spectral width and the decrease in $g$-value are closely related to the SOI.

The temperature dependences of ESR spectrum in K$_n$/K$_{12}$-A at $n = 3.2$ and $n = 7.2$ are shown in Figs.~\ref{fig:ESR-spect}(a) and \ref{fig:ESR-spect}(b), respectively.  Temperature is indicated for each curve.  The magnetic field is normalized at 9.7 GHz to have a common $g$-value which is indicated at the top of the figure.  Spectra at $T = 6.4$ and 4 K in Fig.~\ref{fig:ESR-spect}(a) are in the condition of ferromagnetic resonance.  The $g$-value estimated from the central field shifts to higher and lower sides with a decrease in temperature at paramagnetic temperatures in Figs.~\ref{fig:ESR-spect}(a) and \ref{fig:ESR-spect}(b), respectively.  The spectral width decreases with a decrease in temperature in Fig.~\ref{fig:ESR-spect}(b).

The temperature dependences of $g$-value estimated automatically from the central field of the EPR (or ESR) spectrum and the FWHM of the spectrum are plotted in Figs.~\ref{fig:(K_KA)g-T}(a) and \ref{fig:(K_KA)g-T}(b), respectively, at various values of $n$ in K$_n$/K$_{12}$-A.  The horizontal dotted line in Fig.~\ref{fig:(K_KA)g-T}(a) is the $g$-value of bulk K metal.  

In Fig.~\ref{fig:(K_KA)g-T}(a), the $g$-value for $n<2$ is nearly temperature independent and is near the value of bulk K metal.  The $g$-value at 300 K suddenly decreases for $n>2$ and decreases further with $n$ \cite{Nakano2002ESR}.  The $g$-value decreases markedly with a decrease in temperature for $2<n<5$ above the Curie temperature.    Below the Curie temperature, the $g$-value estimated automatically from the central field of the ESR spectrum quickly increases in Fig.~\ref{fig:(K_KA)g-T}(a). They are in the condition of ferromagnetic resonance, and are not the $g$-value in the original meaning.  We don't analyze them here because of the complexity.  The $g$-values at $n=5.6$ and 6.4 are nearly temperature independent, but  the $g$-value at $n=7.2$ increases markedly with a decrease in temperature. The $n$ dependences of $g$-value at 10 and 300 K are plotted in Fig.~\ref{fig:g-10K-n}. A sudden decrease at $n > 2$ is seen for 10 K. A remarkable decrease is seen at $n = 7.2$ for 300 K.  These behaviors are discussed in terms of the degeneracy of 1$p$ states for $n > 2$ and also the dynamical property of 1$p$ states at higher temperatures in Section~\ref{sec:Electronic1p}.

In Fig.~\ref{fig:(K_KA)g-T}(b), the spectral width for $n<2$ is rather narrow  and nearly temperature independent.  The spectral width increases for $n>2$.  The spectral width for $2<n<5$ is nearly temperature independent, but that  at $n=5.6$ and 6.4 first decreases and then increases with a decrease in temperature. The spectral width at $n=7.2$ markedly decreases with a decrease in temperature.   The increase in FWHM is closely related to the decrease in $g$-value, except for when ferromagnetic resonance occurs.  

If $s=1$ magnetic moments are distributed in a sample as proposed in the ferrimagnetic model~\cite{Maniwa1999}, an EPR signal corresponding to $\Delta s_z = \pm 2$ is expected at half of the resonance magnetic field of $\Delta s_z = \pm 1$.  In the present experiment, however, any such signal was below the detection limit, indicating that few $s=1$ magnetic moments were distributed.

%%%%%%%%%%%%%%%%%%%%%%%%%%%%%%%%%%%
\subsection{\label{sec:Electrical}Electrical Properties}

The electrical conductivity and its temperature dependence give important information on the carriers in a solid, especially in correlated polaron systems.   In K$_n$/K$_{12}$-A, finite optical gaps have been observed in infrared absorption spectra, and an insulating property has been expected indirectly \cite{Nakano1999-KA}.  In the present study, the electrical resistivity was measured directly and the activation energy was estimated from the temperature dependence.  

\begin{figure}[ht]
\resizebox*{6.5cm}{!}{\includegraphics{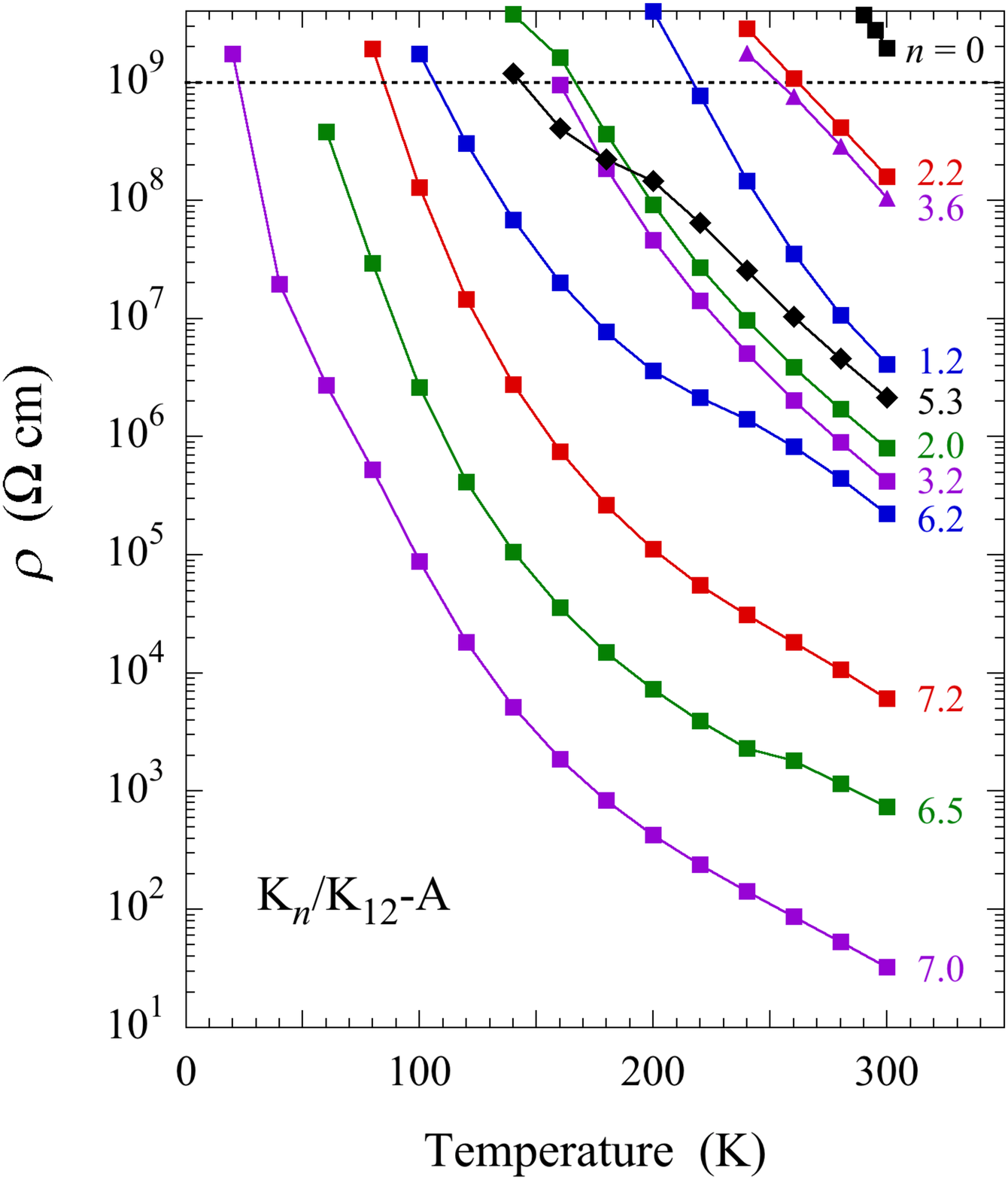}}
\caption{\label{fig:rho-T}(Color online) Temperature dependence of the electrical resistivity $\rho$ in K$_n$/K$_{12}$-A at various values of $n$.  The value of $n$ is indicated for each curve. Values for $\rho \gtrapprox$ 10$^9$ $\Omega\,$cm are unreliable.}
\end{figure}

\begin{figure}[ht]
\resizebox*{6.5cm}{!}{\includegraphics{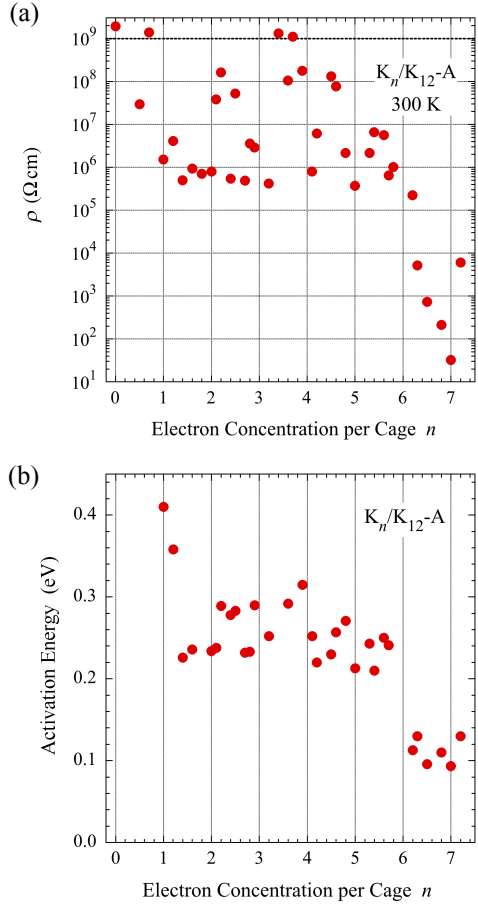}}
\caption{\label{fig:rho-300K}(Color online) (a) Electrical resistivity $\rho$ at 300 K in K$_n$/K$_{12}$-A at various values of $n$.   Values for $\rho \gtrapprox$ 10$^9$ $\Omega\,$cm are unreliable. (b) Roughly estimated thermal activation energy of the electrical conductivity $\sigma=1/\rho$ in K$_n$/K$_{12}$-A at various values of $n$. Data for $\rho \gtrapprox$ 10$^9$ $\Omega\,$cm at 300 K are not shown.}
\end{figure}

Figure~\ref{fig:rho-T} shows the temperature dependence of the electrical resistivity $\rho$ in K$_n$/K$_{12}$-A at various values of $n$.   The unloaded sample ($n = 0$) exhibits a very high $\rho$ around RT, because of ionic conduction.  By the loading of K metal at $n = 1.2$ and 2.0, $\rho$ becomes $\approx\,$$4 \times 10^6$ and $\approx\,$$8 \times 10^5$ $\Omega\,$cm at 300 K, respectively.  With decreasing temperature, $\rho$ increases rapidly.  When $n$ increases to 3.6, $\rho$ becomes $\approx 1 \times 10^8$ $\Omega\,$cm at 300 K.  This value is about two orders larger than that at $n = 3.2$.  With increasing $n$ to 7.0, $\rho$ again decreases to  $\approx\,$$30$ $\Omega\,$cm at 300 K.  At the maximum loading density $n = 7.2$, $\rho$ increases to $\approx\,$$6 \times 10^3$ $\Omega\,$cm at 300 K.  Weakly anomalous behavior is seen at $n = 5.3$ and 6.2 around 200 K.   

The $n$-dependence of $\rho$ at 300 K is plotted in Fig.~\ref{fig:rho-300K}(a). The values for $\rho \gtrapprox$ 10$^9$ $\Omega\,$cm are unreliable. $\rho$ for $n \lessapprox 6$ is scattered over several orders at $\rho \gtrapprox 10^6$ $\Omega\,$cm, and several peaks are seen.  Very high values at $\gtrapprox\,$10$^8$ are checked by the change of compression force, and are confirmed to be reliable.  The electrical resistivity clearly decreases for $n \gtrapprox 6$ and has a shoulder at $n = 7.2$. 

The thermal activation energy of the electrical conductivity $\sigma=1/\rho$ was roughly estimated, shown in Fig.~\ref{fig:rho-300K}(b).   The activation energies for the samples exhibiting $\rho \gtrapprox$ 10$^9$ $\Omega\,$cm at 300 K are not plotted, because of the unreliability of the measurements. The activation energies for $1.4 \lessapprox n  \lessapprox 6$ are $\approx\,$0.25 eV, and those for $n \gtrapprox 6$ are $\approx\,$0.1 eV.  

The conductivity $\sigma$ for different types of carriers is given by 
\begin{eqnarray}
\sigma = \sum\limits_j {e{\mu _j}{N_j}},
\label{eq:conductivity}
\end{eqnarray}
where $e$, $\mu _j$, and $N_j$ are the elementary electric charge, the $j$-th carrier mobility, and the number density of $j$-th carriers, respectively.  There are two limiting models for the electrical conductivity following the Arrhenius law \cite{Ziese1998}.  In the band gap model with nearly temperature independent mobility, the conductivity is proportional to the number density of free carriers, and is expressed by the Arrhenius law. The gap energy is estimated to be twice the thermal activation energy. In the small polaron hopping model, the Arrhenius law can be applied approximately to the temperature dependence of the mobility, where the thermal activation energy is related to the polaron formation energy, etc. The disorder makes a finite contribution to the electrical conductivity.

If ${\mu _j}{N_j}$ has the thermal activation energy $E_j$, $\sigma$ is given by
\begin{eqnarray}
\sigma  = \sum\limits_j {{\sigma_{j0}}\exp \left( { - \frac{E_j}{{{k_{\rm{B}}}T}}} \right)},
\label{eq:conductivity2}
\end{eqnarray}
where $\sigma_{j0}$ is the conductivity of the $j$-th carriers at infinite temperature.  The value of $\exp (-{E_j}/{k_{\rm{B}}}T)$ for $E_j =$ 0.1 eV is about 150 times larger than that for $E_j =$ 0.25 eV at 300 K.  The two orders of decrease in $\rho$ at $n \approx 6.5$ in Fig.~\ref{fig:rho-300K}(a) can be explained mainly by the decrease in the activation energy. The further decrease in $\rho$ at $n=7.0$ can be explained by the increase in $\sigma_{j0}$ for the relevant carriers.

%%%%%%%%%%%%%%%%%%%%%%%%%%%%%%%%%%%
\section{Discussions}

\subsection{\label{sec:tUSn}$t$-$U$-$S$-$n$ Model of Correlated Polaron System}

If the $s$-electrons are well localized quantum mechanically in the zeolite cages with narrow windows, the tight-binding approximation can be applied to the $s$-electrons of the alkali metals in the zeolites \cite{Arita2004, Aoki2004}.  A narrow energy band of $s$-electrons with a strong electron correlation is expected in K$_n$/K$_{12}$-A, because of  the large mutual Coulomb repulsion energy within the $\alpha$-cages and narrow 8R windows \cite{Arita2004}.  Furthermore, the $s$-electrons interact with the displacement of the alkali cations distributed in the cages. Hence, the $s$-electrons have an electron--phonon deformation-potential interaction as well as the electron correlation.  

In order to take an overview of the electronic properties of alkali metals in zeolites, it is effective to introduce the coarse-grained parameters of the correlated polaron system given by the so-called Holstein--Hubbard Hamiltonian \cite{Nakano2017-APX, Anderson1975, Shinozuka1987}:  
\begin{eqnarray} 
\begin{aligned}
H &= -\sum\limits_{i,j,\sigma } {{t_{ij}}a_{i\sigma }^\dag {a_{j\sigma }}}  + U\sum\limits_i {{n_{i \uparrow }}{n_{i \downarrow }}} \\
&+ \sum\limits_i {\left( {\frac{{P_i^2}}{{2m}} + \frac{1}{2}m{\omega ^2}Q_i^2} \right)}  - \lambda \sum\limits_i {{Q_i}\left( {{n_{i \uparrow }} + {n_{i \downarrow }}} \right),}
\end{aligned}
\label{HHH}
\end{eqnarray}
where $a_{i\sigma }$ and $a_{i\sigma }^\dag$  are the annihilation and creation operators of an electron with spin $\sigma$ at the $i$-th site, ${n_{i\sigma }} = a_{i\sigma }^\dag {a_{i\sigma }}$, $t_{ij}$ is the electron transfer energy between the $i$-th and the $j$-th sites,  $U$ is the on-site Coulomb repulsion energy (the Hubbard $U$), and $Q_i$ and $P_i$ are the lattice distortion and the conjugated momentum at the $i$-th site, respectively.  Localized phonons (Einstein phonons) with mass $m$ and frequency $\omega$ are assumed in the third term.   In the last term, the on-site electron--phonon interaction is introduced by assuming the site diagonal coupling constant $\lambda$. Here, we define the lattice relaxation energy $S$ as \cite{Shinozuka1987}
\begin{eqnarray} 
S = \frac{{{\lambda ^2}}}{{m{\omega ^2}}}.
\end{eqnarray}
If we consider the electron transfer between the nearest neighbor sites $\left\langle {i,j} \right\rangle $ only, the first term of the right-hand side of Eq.~(\ref{HHH}) can be written as
\begin{eqnarray} 
- t\sum\limits_{\left\langle {i,j} \right\rangle ,\sigma } {a_{i\sigma }^\dag {a_{j\sigma }}},
\end{eqnarray}
where $t$ $(> 0)$ is the transfer energy of an electron to the nearest neighbor site. The $t$-$U$-$S$-$n$ model of a correlated polaron system is introduced based on the above model, where $n$ is the average number of electrons per site.  

\begin{figure}[ht]
\resizebox*{7.5cm}{!}{\includegraphics{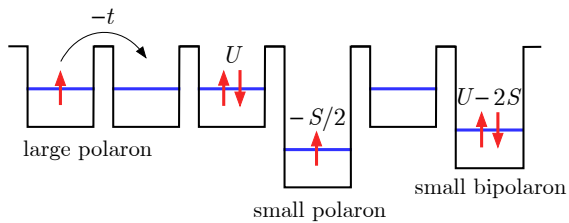}}
\caption{\label{fig:HubbardModel}(Color online) Schematic illustration of the Holstein--Hubbard model.  The red arrows indicate the spins of the electrons. If $t$ is large enough, large polarons are stabilized as free carriers. At small $t$, small polarons with energy $-S/2$ and small bipolarons with energy $U-2S$ are stabilized in cases of $U > S$ and $U < S$, respectively.}
\end{figure}

A schematic illustration of the Holstein--Hubbard model is given in Fig.~\ref{fig:HubbardModel}.  The red arrows indicate the spins of the electrons. If $t$ is large enough, large polarons are stabilized as free carriers.
At small $t$, there are two cases. If $U > S$, small polarons with energy $-S/2$ are stabilized. On the other hand, if $U < S$, small bipolarons with energy $U-2S$ are stabilized.

In zeolites, $t$ is introduced between adjoining cages through windows. The energy band width $2B$ is given by $2B = 2ht$, where $h$ is the number of nearest neighbor sites ($h=6$ for an $\alpha$-cage in zeolite A). The energy of the band bottom is located at $-B$. Two $s$-electrons in the same cage have a Coulomb repulsion energy $U$. The unscreened $U$ is estimated to be $\approx\,$4 eV for an $\alpha$-cage with an inside diameter of $\approx\,$11 \AA{} \cite{Nakano2017-APX, Nozue1993-KA, Arita2004}.  A qualitative interpretation has been given for various properties of alkali metals in different zeolites by the $t$-$U$-$S$-$n$ model \cite{Nakano2017-APX}.  

At lower loading densities in zeolites, $t$ is relatively small because the $s$-electrons occupy the lower quantum states of clusters, and the electron--phonon interaction $S$ dominates the system.  The effective value of $t$ for the energy band near the Fermi energy is expected to increase with $n$, because the $s$-electrons occupy higher quantum states of the clusters \cite{Nakano2017-APX}.

%%%%%%%%%%%%%%%%%%%%%%%%%%%%%%%%%%%
\subsection{\label{sec:ElectronicStates1}K Clusters for $n<2$}

According to the SQW model in Fig.~\ref{fig:Cluster}, 1$s$-states of K clusters are partly occupied with $s$-electrons for $n<2$. At $n=1.0$ and 1.6 in K$_n$/K$_{12}$-A, only one optical reflection peak at 1.3 eV is observed at $\approx78$ K in Figs.~\ref{fig:Ref-LT} and \ref{fig:Ref1.6}(a).  The 1.3-eV band is consistently assigned to the 1$s$--1$p$ excitation in the SQW model indicated in Fig.~\ref{fig:Cluster}.  Hence, $s$-electrons are distributed at the 1$s$-states of clusters in $\alpha$-cages at low temperatures.  The magnetism is nearly nonmagnetic, as explained in  Section~\ref{sec:Magnetic}.  The electrical resistivities are relatively high. The electronic state for $n < 2$ can be assigned to nonmagnetic insulators of small bipolarons of $s$-electrons in 1$s$ states of K clusters.  According to the $t$-$U$-$S$-$n$ model, small bipolarons are formed for $U < S$ with small $t$ and large $S$, and are randomly distributed in zeolite A.  A small number of paramagnetic moments are distributed, because unpaired states of small bipolarons are generated as small polarons with magnetic moments.  The decrease in $g$-values is small, and the ESR widths are narrow at any temperature, because the 1$s$ state does not have angular momentum, as explained in Section~\ref{sec:ESR}.

In a theoretical calculation for $n=1$ with $S=0$ in the assumed structure of K$_n$/K$_{12}$-A, the 1$s$ energy band width has been estimated to be $\approx\,$0.1 eV \cite{Arita2004}, which is very narrow.  The average crystal structure at $n = 0.35$ is assigned to $F{\bar 4}3c$, where structures of K clusters in the adjoining $\alpha$-cages are crystallographically equivalent to each other \cite{Ikeda2004KA}. The real structure, however, has a disorder due to randomly distributed small bipolarons.

%%%%%%%%%%%%%%%%%%%%%%%%%%%%%%%%%%%
\subsection{\label{sec:ElectronicStates2}K Clusters for $n>2$}

According to the SQW model in Fig.~\ref{fig:Cluster}, 1$p$-states of K clusters are partly occupied by $s$-electrons for $n>2$.  The superlattice structure, the optical reflection band at 0.7 eV assigned to $\sigma$-bonding, the decrease in $g$-value, etc., are observed in addition to the ferromagnetic properties.

\subsubsection{\label{sec:Superlattice}Superlattice Structure for $n>2$}

\begin{figure}[ht]
\resizebox*{6.5cm}{!}{\includegraphics{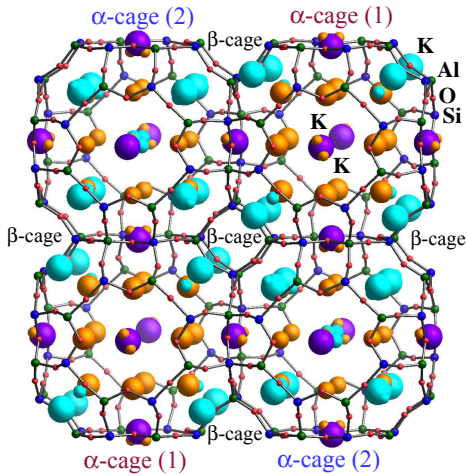}}
\caption{\label{fig:F23}(Color online) Half of the unit cell of the average structure in K$_n$/K$_{12}$-A at $n=4.9$ at 11 K in the space group $F23$, with structure data taken from Ref.~[\onlinecite{Ikeda2004KA}]. The illustrated diameter of each K cation is proportional to the square root of its occupancy.}
\end{figure}

Superlattice reflections observed for $n>2$ in diffraction experiments indicate that the structures of K clusters in adjoining $\alpha$-cages are not crystallographically equivalent to each other. The average crystal structure for $n>2$ is proposed to be $F23$ \cite{Ikeda2000KA, Ikeda2004KA}.  Half of the unit cell structure of K$_n$/K$_{12}$-A at $n=4.9$ at 11 K in the space group $F23$ is illustrated in Fig.~\ref{fig:F23}, where the structure data are taken from Ref.~[\onlinecite{Ikeda2004KA}].  Unfortunately, the $s$-electron wave function was not detected in the X-ray diffraction experiment, because of the extremely low electron density. In the figure, the K cations are indicated by light blue, violet, and orange spheres.   The illustrated diameter of each K cation is in proportion to the square root of the occupancy, where the cross section of the largest sphere corresponds to the occupancy of $\approx\,$80\%. The real occupations of the K cations within the distance 2.8 \AA{} are exclusive, because the radius of a K cation is $\approx\,$1.4 \AA{}. The light blue K cations sit on the axes of the three-fold rotational symmetry.   The violet K cations are located at the centers of the 8Rs.   The orange K cations have 12 equivalent sites and are mainly located beside 4R$_1$s. The orange K cations have relatively low occupancies.  It is suggested that the symmetries of the real structures are lower than the space group $F23$, especially for the orange K cations. 

K cations near the centers of 6Rs (light blue color) are mainly located outside an $\alpha$-cage (1) (namely inside a $\beta$-cage) and inside an $\alpha$-cage (2), indicating that the Coulomb potential of K cations for $s$-electrons in an $\alpha$-cage (2) is deeper than that in an $\alpha$-cage (1).  Hence, the number of $s$-electrons in an $\alpha$-cage (2) is expected to be larger than that in an $\alpha$-cage (1). The difference in the numbers of K cations between adjoining $\alpha$-cages increases at $n \approx 7$ \cite{Ikeda2004KA}. The optical spectrum, however, is insensitive to the difference in the potential depth, because the excitation energy is determined by the size of an $\alpha$-cage, as shown in Fig.~\ref{fig:Cluster}.  On the other hand, the magnetic properties are quite sensitive to the difference in the electronic configuration. 

$\sigma$-bonding between 1$p$ states of K clusters in adjoining $\alpha$-cages is expected for $n > 2$, as discussed in Sections \ref{sec:sigma} and \ref{sec:Electronic1p}.   Why the superlattice structure is generated for $n > 2$ is an interesting question. The most plausible reason is that the alternate order of the K cluster structure is stabilized by the asymmetric $\sigma$-bondings between adjoining K clusters, as explained in the next section. 

%%%%%%%%%%%%%%%%%%%%%%%%%%%%%%%%%%%
\subsubsection{\label{sec:sigma}Asymmetric $\sigma$-Bonding}

In the $t$-$U$-$S$-$n$ model, small $S$ (large $U$ and/or large $t$) regions generate a rather homogeneous distribution of electrons, such as in Mott insulators and metals, because $U$ separates electrons and $t$ contributes to the motion of electrons over many sites.  On the other hand, a large $S$ region provides an inhomogeneous distribution of electrons, such as small bipolarons in which two electrons are self-trapped at the same site, because $S$ combines repelling electrons. The problem of electronic configuration in adjoining sites is equivalent to \textit{two-site and two-electron systems} \cite{Toyozawa1981}.  The region at $S = 0$ has two limits: the molecular orbital limit for $t \gg U$ and the Heitler--London schemes (valence bond theory) for $t \ll U$.  In both limits, the ground states of two adjoining sites are symmetric and in the spin-singlet.  In the molecular orbital limit, the $\sigma$-bonding and anti-$\sigma$-bonding states between two adjoining sites are formed. Two electrons occupy the $\sigma$-bonding state. In the valence bond theory, two electrons occupy two sites separately, and the first excited state is spin-triplet with a small excitation energy $\approx\,$$4t^{2}/U$ (magnetic excitation). The charge transferred state, where one site has more electrons and the other has less electrons (asymmetric), is formed in the spin-singlet state as the second excited state with excitation energy $\approx\,$$U$.  If we assume a large $S$ ($S > t$), the energy of the above charge transferred state decreases as $\approx U - 2S$, and becomes the lowest state for $U < S$, indicating that the charge transferred asymmetric state has an energy lower than that of the symmetric state (the replacement of the ground states) \cite{Toyozawa1981}.  This mechanism is assigned to charge transfer instability with a change of structure.  

In the $t$-$U$-$S$-$n$ model, two-cage type small bipolarons cannot be stabilized without an additional local structure.  If the $s$-electrons have a large $t'$ ($>t$) between adjoining cages due to the additional local structure in the arrangement of the K cations, a two-cage type small bipolaron state can be stabilized there as asymmetric $\sigma$-bonding.  The optical excitation is allowed from the $\sigma$-bonding to the anti-$\sigma$-bonding at an excitation energy of $\approx\,$$2t'$. The asymmetric $\sigma$-bonding between 1$p$ states in adjoining cages through 8Rs is theoretically expected in one of the model structures at $n = 4$ \cite{Nohara2009}.  The additional local structure can be provided by the distribution of orange K cations beside the 4R$_1$s shown in Fig.~\ref{fig:F23}.

%%%%%%%%%%%%%%%%%%%%%%%%%%%%%%%%%%%
\subsubsection{\label{sec:Electronic1p}Electronic Properties for $n>2$ and Hypothetical Models}

In the reflection spectra at $\approx\,$78 K in Fig.~\ref{fig:Ref-LT}, a peak at 0.7 eV appears for $2 < n < 6$. This band is assigned to the $s$-electron excitation from a 1$p$-related state to a higher state.  The 1$p$--1$d$ excitation energy in the SQW potential model in Fig.~\ref{fig:Cluster} is expected to be around 1.5 eV, and the corresponding band is observed in Fig.~\ref{fig:Ref-LT}.  The energy of 0.7 eV is too low to be assigned to the 1$p$--1$d$ excitation.  On the other hand, asymmetric $\sigma$-bonding between adjoining cages is expected, as discussed in Section~\ref{sec:sigma}. The most plausible $\sigma$-bondings are along the $x$-, $y$-, and $z$-directions between the $1p_x$, $1p_y$, and $1p_z$ orbitals in adjoining $\alpha$-cages, respectively.  The energy of excitation from the $\sigma$-bonding to the anti-$\sigma$-bonding is expected to be $\approx\,$$2t'$. Hence, the 0.7 eV band can possibly be assigned to the excitation of the $\sigma$-bonding of 1$p$ states to its anti-$\sigma$-bonding.

\begin{figure}[ht]
\resizebox*{7.5cm}{!}{\includegraphics{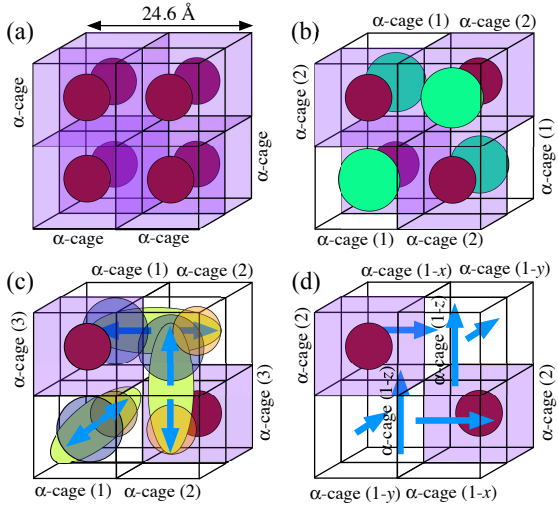}}
\caption{\label{fig:Cube}(Color online) Schematic illustration of the ordering of K clusters in cubic structures.  (a) K clusters in the $F{\bar 4}3c$ structure expected at $n = 0.35$. The eight K clusters are equivalent to each other. (b)  K clusters in the $F23$ structure expected for $n > 2$. The two types of K clusters are arrayed alternately in $\alpha$-cages (1) and (2). (c) Hypothetical cubic model, differing slightly from the $F23$ structure with three pairs of K clusters in adjoining $\alpha$-cages (1) and (2), and two K clusters in $\alpha$-cages (3) in the unit cell. (d) Hypothetical cubic model differing slightly from the $F23$ structure with an orbital orthogonalization of $1p_x,$ $1p_y,$ and $1p_z$ holes in $\alpha$-cages (1-$x$), (1-$y$), and (1-$z$), respectively.}
\end{figure}

The original zeolite A has an $Fm{\bar 3}c$ structure with eight equivalent $\alpha$-cages in the unit cell, as shown in Fig.~\ref{fig:LTA}. For $n = 0.35$, the $F{\bar 4}3c$ structure is proposed  \cite{Ikeda2004KA}, and the average structures of the eight K clusters are equivalent to each other, as shown schematically in Fig.~\ref{fig:Cube}(a), where the large spheres in cubes are schematic images of $s$-electron wave functions which are not detected directly from diffraction experiments.  For $n> 2$, the $F23$ structure is proposed \cite{Ikeda2000KA, Ikeda2004KA}.  Two types of K clusters are alternately ordered, as shown in Fig.~\ref{fig:Cube}(b), where $\alpha$-cages (1) and (2) accommodate distinct K clusters such as those in Fig.~\ref{fig:F23}. The pairing of K clusters by the $\sigma$-bonding between adjoining cages, however, is difficult in the $F23$ structure. If we assume four $\sigma$-bondings in the unit cell, for example, the symmetry cannot be cubic.  Hence, we assume a hypothetical model with three $\sigma$-bondings in the cubic unit cell, which is slightly different from the $F23$ structure, as follows.

We assume three types of $\alpha$-cages (1), (2), and (3), as shown in Fig.~\ref{fig:Cube}(c), where a $(-1/4, -1/4, 1/4)$ shift of the unit cell is required for a cubic lattice. The center of the unit cell is set at the center of $\alpha$-cage (3). There are three pairs of K clusters in adjoining $\alpha$-cages (1) and (2), and two other K clusters in $\alpha$-cages (3). The $\sigma$-bondings constructed of $1p_x, 1p_y,$ and $1p_z$ orbitals in adjoining $\alpha$-cages are illustrated by spheroids with arrows, where the directions of the arrows show the signs of the $1p_x,$ $1p_y,$ and $1p_z$ orbitals.  Two opposite arrows in a spheroid indicate $\sigma$-bonding. The direction of an arrow, however, is not detected from diffraction experiments.

\begin{figure}[ht]
\resizebox*{7.5cm}{!}{\includegraphics{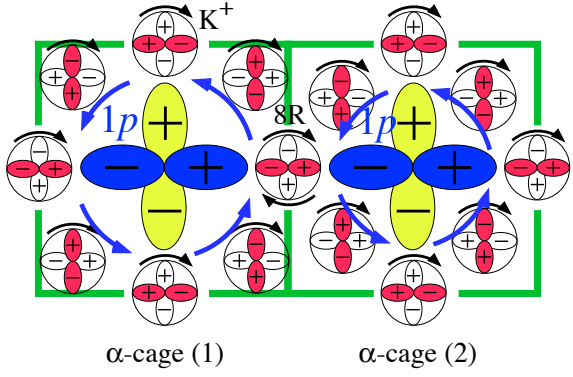}}
\caption{\label{fig:DM-Rashba}(Color online) Schematic illustration of an example of the Rashba type DM interaction between adjoining $1p$ orbitals at $\alpha$-cages (1) and (2).  Adjoining $1p$ orbitals share K cation at the center of 8R.  If there is no inversion symmetry at the center of 8R, a finite DM interaction is expected.  }
\end{figure}

As explained in Section~\ref{sec:level1}, a huge spin-orbit interaction is expected at  $1p$ orbitals due to the Rashba mechanism \cite{Nakano2017-APX, Nakano2007}. Here, we propose an enhancement effect of the DM interaction  between adjoining $1p$ orbitals by the extension of the Rashba mechanism. An example is illustrated schematically in Fig.~\ref{fig:DM-Rashba}, where adjoining $1p$ orbitals are located at $\alpha$-cages (1) and (2).  Counterclockwise rotation of a $1p$ electron induces clockwise rotations of core electrons of K cations at the equator in each cage.  Adjoining $1p$ orbitals share K cation at the center of 8R.  If there is no inversion symmetry at the center of 8R, a finite DM interaction is expected between antiferromagnetic spins at adjoining $1p$ orbitals.  K cations at other centers of 8Rs induce similarly the DM interaction between other $1p$ orbitals, and magnetic domains with the spontaneous magnetization are stabilized  for ferromagnetism. Zeolite frameworks are well ordered, but K cations have disorders in their positions and occupancies depending on the loading density $n$. A thermal fluctuation of cation positions are expected even at low temperatures because of a weak interaction with framework. The basic magnetic structure of present material is a disordered antiferromagnetism of the Mott insulator, and related DM interactions are expected to be disordered as well.

At $n = 7.2$, the Curie law has been observed, although spin $s = 1/2$ magnetic moments are distributed in $\approx 80$\% of $\alpha$-cages, as explained in Section~\ref{sec:Magnetic}.  The Curie law indicates that adjacent clusters have no magnetic interaction \cite{Nakano1999-KA}. This result is explained by the mutually orthogonalized nondegenerate holes in 1$p$-states \cite{Nakano1999-KA, Nakano2002ESR}. In order to realize the orbital orthogonalization structure, $1p_x,$ $1p_y,$ and $1p_z$ holes are assumed in $\alpha$-cages (1-$x$), (1-$y$), and (1-$z$), respectively, as shown Fig.~\ref{fig:Cube}(d).  Adjoining 1$p$-orbitals (holes) are orthogonalized with each other, as illustrated by arrows.  Full occupancy clusters are assumed at $\alpha$-cages (2).

\begin{figure}[ht]
\resizebox*{8.0cm}{!}{\includegraphics{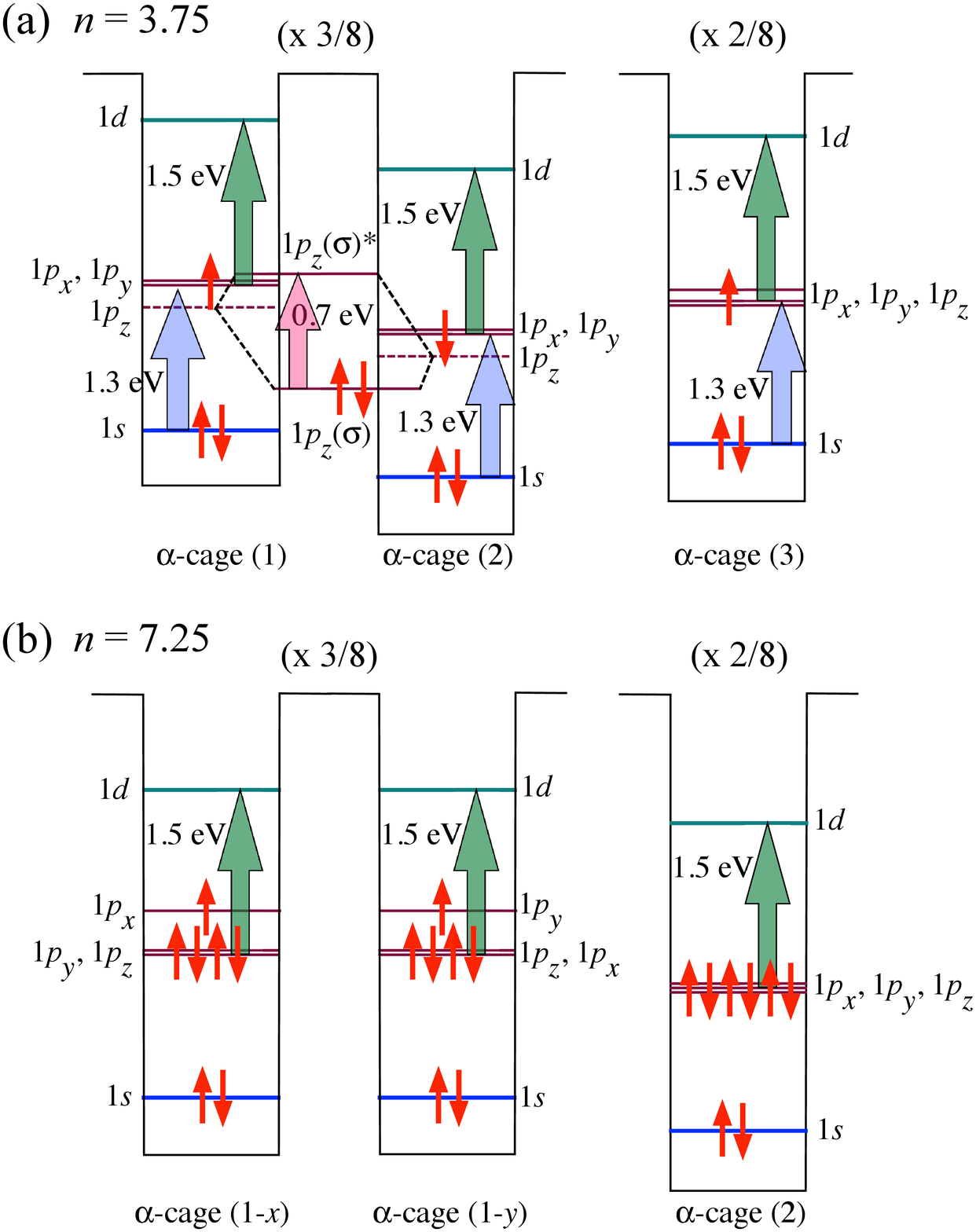}}
\caption{\label{fig:sigma}(Color online) Configurations of $s$-electrons (a) and (b) according to the hypothetical cubic models in Figs.~\ref{fig:Cube}(c) and \ref{fig:Cube}(d), respectively.  The red arrows indicate the spins of the $s$-electrons.  Corresponding optical excitations are indicated by large arrows with respective excitation energies. In configuration (a), the average $s$-electron concentration per cage and the average population of magnetic moments are 3.75 and 100\%, respectively. In configuration (b), the average $s$-electron concentration per cage and the average population of magnetic moments are 7.25 and 75\%, respectively.}
\end{figure}

Possible examples of the configurations of $s$-electrons are illustrated in Figs.~\ref{fig:sigma}(a) and \ref{fig:sigma}(b), from a consideration of the hypothetical cubic model shown in Figs.~\ref{fig:Cube}(c) and \ref{fig:Cube}(d), respectively, where the red arrows indicate the spins of the $s$-electrons.  Corresponding optical excitations are indicated by large arrows with respective excitation energies. In the two figures, the average $s$-electron concentrations per cage are 3.75 and 7.25, respectively, and the average populations of magnetic moments are 100 and 75\%, respectively. 

In the electronic configuration in Fig.~\ref{fig:sigma}(a), the asymmetric $\sigma$-bonding state 1$p_z(\sigma)$ has two electrons in the spin-singlet state.  The 0.7-eV band in the reflection spectra can be explained by the excitation from the $\sigma$-bonding 1$p_z(\sigma)$ to the anti-$\sigma$-bonding 1$p_z(\sigma)$*. An $s$-electron occupies each 1$p_x$-1$p_y$ degenerate state in adjoining $\alpha$-cages (1) and (2). One more $s$-electron occupies a 1$p$ state in $\alpha$-cage (3), where triply degenerate 1$p$ states in a real system are difficult, because of the difficulty of the high symmetry distribution of cations. There are three pairs of $\alpha$-cages (1) and (2) and two $\alpha$-cages in the unit cell with eight $\alpha$-cages.  In this model, $s$-electrons in degenerate 1$p$ states have a large DM interaction, according to the theory of the DM interaction \cite{Tachiki1968}.  The DM interaction plays an important role in the spin-cant model of degenerate 1$p$ states, as explained in Section~\ref{sec:Cluster}. The ferromagnetic properties at $n \approx 3.6$ in Section~\ref{sec:Magnetic} can be explained by the electronic model in Fig.~\ref{fig:sigma}(a).

With a decrease in $n$ from 3.75 in Fig.~\ref{fig:sigma}(a), some disorder and a decrease in the population of magnetic moments are expected, but detailed structure models are difficult to be determined.   The Curie constant increases with $n$ for $2 < n < 3.5$ in Fig.~\ref{fig:CurieConst}. In Figs.~\ref{fig:MT-1}(a) and \ref{fig:MT-1}(b), the ferromagnetic volume with similar Curie temperature seems to increase with $n$ for $2 < n < 3.5$.  Therefore, the volume of ferromagnetic domains which have an electronic configuration similar to that in Fig.~\ref{fig:sigma}(a) may increase with $n$ for $2 < n < 3.5$.  

On the other hand, with an increase in $n$ from 3.75, $s$-electrons can occupy 1$p_z(\sigma)$* states. The full occupation of $\sigma$-bonding and anti-$\sigma$-bonding states, however, generally causes a dissociation of the $\sigma$-bonding, because there is no energy gain in the total bondings. The disappearance of the 0.7-eV band for $n > 6$ in Fig.~\ref{fig:Ref-LT} can be explained by the dissociation of all the $\sigma$-bondings.  If the dissociation of the $\sigma$-bonding stabilizes the mutually orthogonalized nondegenerate holes in 1$p$-states shown in Fig.~\ref{fig:sigma}(b), the magnetic interactions are disconnected with an increase in $n$, and the Curie temperature decreases with $n$, as observed in Figs.~\ref{fig:MT-2}(a) and \ref{fig:MT-2}(b), and as plotted in Fig.~\ref{fig:Curie}.  Determining detailed local structures of the K clusters is difficult, but different structures are possible at the K cations beside 4R$_1$s with low occupancies in the average structure of $F23$ symmetry in Fig.~\ref{fig:F23}. As the double occupation of $s$-electrons at degenerate 1$p$ states are suppressed by the large $U$, $s = 1$ spin-triplet states with Hund's coupling are scarcely distributed, as mentioned in Section~\ref{sec:ESR}.

At $n = 7.25$ in Fig.~\ref{fig:sigma}(b), six orbital-orthogonalized K-clusters in $\alpha$-cages (1-$x$), (1-$y$), and (1-$z$) have holes in 1$p$-states, and two K-clusters in $\alpha$-cages (2) are fully occupied with $s$-electrons at 1$p$-states.  If 1$p$-holes in adjoining $\alpha$-cages are orthogonalized with each other, a paramagnetic system with the Curie law is realized.  In Figs.~\ref{fig:MT-2}(b) and \ref{fig:Curie}, both $T_{\rm{C}}$ and $T_{\rm{W}}$ are estimated to be 0 at $n = 7.2$, although $\approx\,$80\% of $\alpha$-cages are occupied with magnetic moments of spin $s = 1/2$.   By the pressure loading technique, $n$ can be increased more than the saturation value of 7.2, and a disappearance of magnetic moments has been observed \cite{Nam2007}.  This result is explained by the full occupation of $s$-electrons at 1$p$-states of K clusters in all $\alpha$-cages at $n = 8$.

The decrease in $g$-value for $n > 2$ in comparison with that for $n < 2$ in Fig.~\ref{fig:(K_KA)g-T}(a) clearly indicates that the contribution of orbital angular momentum is large for $n > 2$.  This result coincides well with a model wherein $s$-electrons start to occupy the 1$p$ states with orbital angular momentum for $n> 2$.  The decrease in $g$-value at low temperatures for $2<n<5$ indicates that the contribution of the orbital angular momentum is large at low temperatures.  At high temperatures, thermal vibrations along the adiabatic potential can cancel the effect of degeneracy in 1$p$-states, because the thermally deformed non-degenerate states dominate in the configuration coordinates.  In contrast, the $g$-value at $n = 7.2$ decreases at high temperatures. This result indicates that the non-degenerate 1$p$-states at low temperatures dynamically mix with degenerate 1$p$-states at high temperatures.  These interpretations are consistent with the above-mentioned electronic models on magnetic moments where the degenerate and non-degenerate 1$p$-states are the origins of magnetic moments for $2 < n < 6$ and $n = 7.2$, respectively.

\subsubsection{\label{sec:Mott}Continuous Mott Insulator for $n>2$}

In a $U/t$-dominant system with $S = 0$, the antiferromagnetism of a Mott insulator is stabilized theoretically at the just-half-filled electron density (which corresponds to $n = 1$) \cite{Moriya1980}.    The system, however, changes from an antiferromagnetic insulator to a ferromagnetic metal for $n \ne 1$ \cite{Moriya1980}. If the system has a finite $S$, the antiferromagnetic Mott insulator is stable not only at $n = 1$ but also for $n \ne 1$ under the condition $U > S$ in which the generation of small bipolarons is suppressed \cite{Shinozuka1987}. Namely, a Mott insulator with finite $S$ is robust against deviation from $n = 1$, because small polarons are stabilized. 

In zeolite sodalite, $\beta$-cages are arrayed in a body centered cubic structure by the sharing of 6Rs. Alkali metal clusters in $\beta$-cages display a typical Heisenberg antiferromagnetism of Mott-insulator at $n = 1$, which corresponds to the just-half-filled condition of 1$s$ states in $\beta$-cage clusters \cite{Nakano2017-APX, Srdanov1998, Nakamura2009, Nakano2013JKPS, Nakano2013PRB}.  The antiferromagnetic Mott insulator is robust against deviation from $n = 1$.  The magnetic moments are well localized in $\beta$-cage clusters not only at $n = 1$, because of large $U/t$ and finite $S/t$ for $U > S$. 

In the simplified model structure of K$_{n}$/K$_{12}$-A at $n = 3$, a Mott insulator is theoretically expected for the 1$p$ energy bands, where the band width is estimated to be $\approx\,$0.4 eV \cite{Arita2004}.  In the present experimental results for K$_{n}$/K$_{12}$-A, a Mott insulator phase is observed continuously for $n > 2$.  Finite optical gaps are observed in the infrared absorption spectra at any value of $n$ \cite{Nakano1999-KA}.  A large $U/t$ with an intermediate $S/t$ is realized for 1$p$ states in K$_{n}$/K$_{12}$-A, because an intermediate value of $t$ is expected from the  intermediate size of the 8R-windows of the $\alpha$-cages. Then, a continuous Mott insulator can be realized for $n > 2$.  In contrast to the 1$s$ states for $n < 2$, small bipolarons in the 1$p$ states for $n > 2$ are unstable under the condition $U > S$.    As mentioned in Section~\ref{sec:sigma}, two-cage type small bipolarons are stabilized as $\sigma$-bonding by the additional local structure.

In a low-silica X (LSX) zeolite, supercages of FAU with an inside diameter of $\approx\,$13 \AA{} are arrayed in a diamond structure by the sharing of 12-membered rings (12Rs) with an inside diameter of $\approx\,$8 \AA{}. In contrast to K$_{n}$/K$_{12}$-A, K clusters in LSX acquire a metallic property for $n > 6$, because a large $t$ is expected from the large 12R-windows. Ferrimagnetic properties are observed at $n \approx 9$ \cite{Nakano2013K-LSX} and  an itinerant electron ferromagnetism is observed at $n \approx 15$ under pressure loading of the K metal at $\approx\,$0.9 GPa \cite{Araki2019}.

The electrical conductivity in disordered materials generally consists of contributions from the different carriers, as shown in Eqs.~\ref{eq:conductivity} and \ref{eq:conductivity2}, where the $j$-th carriers contribute $e{\mu _j}{N_j}$. In K$_n$/K$_{12}$-A, the small polaron hopping model is applicable for $n > 2$, because polaron conductivity is expected for an intermediate $S/t$.  The electrical resistivity at 300 K changes dramatically in Fig.~\ref{fig:rho-300K}(a), but the activation energy changes only moderately in Fig.~\ref{fig:rho-300K}(b).  A Mott insulator has a high resistivity at the just-half-filled condition in an ordered system. At the deviation from the just-half-filled condition, disorder may be introduced, and an excess or a deficit of $s$-electrons provides polarons with a finite activation energy.  A small difference in $n$ can change the resistivity dramatically, but the activation energy of polarons is similar at the deviation from the just-half-filled condition.  In Fig.~\ref{fig:rho-300K}(b), similar activation energies of $\approx\,$0.25 eV for $2 \lessapprox n < 6$ are seen, and may be provided by the model in Fig.~\ref{fig:sigma}(a) with an excess or deficit of $s$-electrons.  Similar activation energies of $\approx\,$0.1 eV at $n \approx 7$ in Fig.~\ref{fig:rho-300K}(b) may be provided by the model in Fig.~\ref{fig:sigma}(b)  with a deficit of $s$-electrons.

%%%%%%%%%%%%%%%%%%%%%%%%%%
\section{Summary}

We measured the detailed electronic properties of K$_n$/K$_{12}$-A, and discussed the results with reference to past experimental results based on a $t$-$U$-$S$-$n$ correlated polaron system. The electronic properties vary clearly depending on $n$.  The properties for $n < 2$ are consistent with nonmagnetic insulators, and are explained by the formation of small bipolarons, where the electron--phonon interaction dominates the system ($U < S$).  The properties for $n > 2$ are those of magnetic insulators, where the electron correlation dominates the system ($U > S$).  The properties for $n < 2$ and $n > 2$ can be explained by $s$-electrons in the 1$s$ and the 1$p$ quantum states of K clusters, respectively.  Ferromagnetic properties are observed for $n > 2$, and the Curie temperature has a peak at $n \approx 3.6$, and becomes 0 K at $n = 7.2$. An optical reflection band at 0.7 eV is clearly observed for $2 < n < 6$ at low temperatures, and is assigned to the excitation of the $\sigma$-bonding of 1$p$-states in adjoining $\alpha$-cages.  To explain the various ferromagnetic properties, hypothetical electronic models with $n = 3.75$ and 7.25 are proposed including a superlattice structure. The electronic model with $n = 3.75$ involves $\sigma$-bonding. The electronic model with $n = 7.25$ involves the orbital orthogonality of 1$p$-hole states and the dissociation of $\sigma$-bondings. The ferromagnetic properties for $2 < n \lessapprox 3.6$ can be explained by ferromagnetic domains based on the electronic model with $n = 3.75$.  The ferromagnetic properties for $3.6 \lessapprox n < 7.2$ are explained by a contribution from the orbital orthogonality in the electronic model with $n = 7.25$. The paramagnetic state at $n = 7.2$ can be explained by the electronic model with $n = 7.25$. A Mott insulator is continuously observed for $n > 2$, and can be explained by correlated polarons with a large $U$ and an intermediate $S$, where $U > S$.

%%%%%%%%%%%%%%%
\begin{acknowledgments}
We are deeply grateful to Profs./Drs. R. Arita, Y. Nohara, K. Nakamura, H. Aoki, and T. Ikeda for their theoretical and experimental studies and contributions.   We also thank Mr. S. Tamiya for his chemical analysis.  This work was supported by a Grant-in-Aid for Scientific Research on Priority Areas (No. JP19051009), Grants-in-Aid for Scientific Research (A) (No. JP24244059 and No. JP13304027) and (C) (No. JP26400334), Grant-in-Aid for Creative Scientific Research ``New Phases of Matter in Multidisciplinary Approaches" (No. JP15GS0213), Global COE Program ``Core Research and Engineering of Advanced Materials-Interdisciplinary Education Center for Materials Science'' (G10), the 21st Century COE Program ``Towards a new basic science: depth and synthesis'' (G17), MEXT Japan, and also by Core Research for Evolutional Science and Technology (CREST)  ``New Arrayed Clusters in Microporous Materials: Syntheses, Structures and Physical Properties'' 7-6, JST.

\end{acknowledgments}

%\nocite{*}


\begin{thebibliography}{10} 

\bibitem{Nakano2017-APX}
T.~Nakano and Y.~Nozue, Adv. Phys.: X {\bf2}, 254 (2017).
% Electrons of alkali metals in regular nanospaces of zeolites

\bibitem{Nakano2013ICMInv}
T.~Nakano, D.~T.~Hanh, Y.~Nozue, N.~H.~Nam, T.~C.~Duan, and S.~Araki, J. Korean Phys. Soc. {\bf63}, 699 (2013).
%Exotic Magnetism of s-electron Cluster Arrays: Ferromagnetism, Ferrimagnetism and Antiferromagnetism


%%%  K/K12-A  %%%

\bibitem{Nozue1992-KA}
Y.~Nozue, T.~Kodaira, and T.~Goto, Phys. Rev. Lett. {\bf68} (1992) 3789.
% Ferromagnetism of Potassium Clusters Incorporated into Zeolite LT A

\bibitem{Nozue1993-KA}
Y.~Nozue, T.~Kodaira, S.~Ohwashi, T.~Goto, and O.~Terasaki, Phys. Rev. B {\bf48}, 12253 (1993).
% Ferromagnetic properties of potassium clusters incorporated into zeolite LTA

\bibitem{Kodaira1993-KA}
T.~Kodaira, Y.~Nozue, S.~Ohwashi, T.~Goto, and O.~Terasaki, Phys. Rev. B {\bf48}, 12245 (1993).
% Optical properties of potassium clusters incorporated into zeolite LTA

\bibitem{Armstrong1994} %aaaaaaa
A. R. Armstrong, P. A. Anderson, P. P. Edwards, J. Solid State Chem. {\bf 111}, 178 (1994).
% K-KA neutron

\bibitem{Nakano1999-KA}
T.~Nakano, Y.~Ikemoto, and Y.~Nozue, Eur. Phys. J. D {\bf9}, 505 (1999).
% Ferromagnetism and paramagnetism in potassium clusters incorporated in zeolite LTA


\bibitem{Maniwa1999}
Y.~Maniwa, H.~Kira, F.~Shimizu, and Y.~Murakami, J. Phys. Soc. Jpn. {\bf68}, 2902 (1999).
% Development of Superstructure in Potassium-Loaded Zeolite LTA, Kx-K12Si12Al12O48

\bibitem{Ikeda2000KA}
T.~Ikeda, T.~Kodaira, F.~Izumi, T.~Kamiyama, and K.~Ohshima, Chem. Phys. Lett. {\bf318}, 93 (2000).
% Neutron powder diffraction study of potassium clusters in zeolite K-LTA


\bibitem{Nakano2000MCLC}
T.~Nakano, Y.~Ikemoto, and Y.~Nozue, Mol. Cryst. Liq. Cryst. {\bf341}, 461 (2000).
% Magnetic Properties Near the Ferromagnetic-Nonferromagnetic Phase Boundary in Potassium Clusters Incorporated into Zeolite LTA

\bibitem{Nakano2000}
T.~Nakano, Y.~Ikemoto, and Y.~Nozue, Physica B {\bf281}\&{\bf281}, 688 (2000).
% Loading density dependence of ferromagnetic properties in potassium clusters arrayed in a simple cubic structure in zeolite LTA

\bibitem{Nakano2001-RbA-KA}
T.~Nakano, Y.~Ikemoto, and Y.~Nozue, J. Magn. Magn. Mater. {\bf226}-{\bf230}, 238 (2001).
% Ferromagnetic properties of rubidium clusters in zeolite LTA


\bibitem{Kira2001} %aaaaaaa
H. Kira, H. Tou, Y. Maniwa, and Y. Murakami, J. Magn. Magn. Mater. {\bf226}-{\bf230}, 1095 (2001).
% Magnetic interaction in K-absorbing zeolite LTA

\bibitem{Kira2002} %aaaaaaa
H. Kira, H. Tou, Y. Maniwa, and Y. Murakami, Physica B {\bf312}-{\bf 313}, 789 (2002).
% Magnetic interaction in K-absorbing zeolite LTA


\bibitem{Nakano2002ESR}
T.~Nakano, Y.~Ikemoto, and Y.~Nozue, J. Phys. Soc. Jpn. Suppl. {\bf71}, 199 (2002).
% Electron Spin Resonance Study and Orbital Degeneracy of Potassium Clusters in Zeolite LTA


\bibitem{Ikeda2004KA}
T.~Ikeda, T.~Kodaira, F.~Izumi, T.~Ikeshoji, and K.~Oikawa, J. Phys. Chem. B {\bf108}, 17709 (2004).
% Crystal Structures of Zeolite Linde Type A Incorporating K Clusters: Dependence on the K Atom Loading Density

\bibitem{Nakano2004}
T.~Nakano, D.~Kiniwa, Y.~Ikemoto, and Y.~Nozue, J. Magn. Magn. Mater. {\bf272}-{\bf276}, 114 (2004).
% Spin-cant model of ferromagnetism in potassium clusters incorporated in zeolite LTA

\bibitem{Arita2004}
R.~Arita, T.~Miyake, T.~Kotani, M.~van~Schilfgaarde, T.~Oka, K.~Kuroki, Y.~Nozue, and H.~Aoki, Phys. Rev. B {\bf69}, 195106 (2004).
% Electronic properties of alkali-metal loaded zeolites: Supercrystal Mott insulators

\bibitem{Aoki2004}
H.~Aoki, Appl. Surf. Sci. {\bf237}, 2 (2004).
% Design of electron correlation effects in interfaces and nanostructures

\bibitem{Nakano2007}
T.~Nakano and Y.~Nozue, J. Comput. Methods Sci. Eng. {\bf7}, 443 (2007).
% Orbital degeneracy and magnetic properties of potassium clusters incorporated into nanoporous crystals of zeolite A

\bibitem{Nam2007}
N.~H.~Nam, S.~Araki, H.~Shiraga, S.~Kawasaki, and Y.~Nozue, J. Magn. Magn. Mater. {\bf310}, 1016 (2007).
% Magnetic property of potassium clusters in pressure-doped zeolite A

\bibitem{Nakano2007-HighMag}
T.~Nakano, D.~Kiniwa, A.~Matsuo, K.~Kindo, and Y.~Nozue, J. Magn. Magn. Mater. {\bf310}, e295 (2007).
% Anomalous magnetization of potassium clusters incorporated into zeolite A at high magnetic field

\bibitem{Nakano2009muSR-KA}
T.~Nakano, J.~Matsumoto, T.~C.~Duan, I.~Watanabe, T.~Suzuki, T.~Kawamata, A.~Amato, F.~L.~Pratt, Y.~Nozue, Physica B {\bf404}, 630 (2009).
% Fast muon spin relaxation in ferromagnetism of potassium clusters in zeoliteA



\bibitem{Nohara2009}
Y.~Nohara, K.~Nakamura, and R.~Arita, Phys. Rev. B {\bf80}, 220410(R) (2009).
% Spin density functional study of magnetism in potassium-loaded zeolite A

\bibitem{Nohara2011}
Y.~Nohara, K.~Nakamura, and R.~Arita, J. Phys. Soc. Jpn. {\bf80}, 124705 (2011).
% Ab initio Derivation of Correlated Superatom Model for Potassium Loaded Zeolite A


%%%
\bibitem{IZA}
{\it International Zeolite Association} (IZA), available at http://www.iza-online.org 
%The structure database is available at http://www.iza-structure.org/databases/.




%%% ELSE  Rb/Rb-A  %%% 
\bibitem{Duan2007}
T.C. Duan, T. Nakano, and Y. Nozue, e-J. Surf. Sci. Nanotech. {\bf 5}, 6 (2007).
% Magnetic and Optical Properties of Rb and Cs Clusters Incorporated into Zeolite A

\bibitem{Duan2007b}
T.~C.~Duan, T.~Nakano, and Y.~Nozue, J. Magn. Magn. Mater. {\bf310}, 1013 (2007).
% Evidence for ferromagnetism in rubidium clusters incorporated into zeolite A


%%% SOD %%% 

\bibitem{Srdanov1998}
V.~I.~Srdanov, G.~D.~Stucky, E.~Lippmaa, and G.~Engelhardt, Phys. Rev. Lett. {\bf80}, 2449 (1998).
% Evidence for an Antiferromagnetic Transition in a Zeolite-Supported Cubic Lattice of F Centers

\bibitem{Sankey1998} %bbbbbbb
O.F. Sankey, A.A. Demkov, and T. Lenosky, Phys. Rev. B {\bf 57}, 15129 (1998).
% Electronic structure of black sodalite

\bibitem{Blake1998} %ref3
N. Blake and H. Metiu, J. Chem. Phys. {\bf 109}, 9977 (1998).
% Self-interaction-corrected band structure calculations for intracavity electrons in electro-sodalite

\bibitem{Blake1999} %ref3
N. Blake and H. Metiu, J. Chem. Phys. {\bf 110}, 7457 (1999).
%“The importance of self-interaction and nonlocal exchange corrections to the density functional theory of intracavity electrons in Na-doped sodalities,” 

\bibitem{Heinmaa2000} %bbbbbbb
I. Heinmaa, S. Vija, E. Lippmaa, Chem. Phys. Lett. {\bf 327}, 131 (2000).
% NMR study of antiferromagnetic black sodalite Na AlSiO / 8 4 6

\bibitem{Madsen2001} %bbbbbbb
G. K. H. Madsen, B. B. Iversen, P. Blaha, and K. Schwarz, Phys. Rev. B {\bf 64}, 195102 (2001).
% Electronic structure of the sodium and potassium electrosodalites \UTF{0084}Na\UTF{00D5}K 8\UTF{0084}AlSiO46

\bibitem{Tou2001} %bbbbbbb
H. Tou, Y. Maniwa, K. Mizoguchi, L. Damjanovic, V.I. Srdanov, J. Magn. Magn. Mater. {\bf 226}-{\bf 230}, 1098 (2001).
% NMR studies on antiferromagnetism in alkali-electro-sodalite

\bibitem{Scheuermann2002} %bbbbbb
R. Scheuermann, E. Roduner, G. Engelhardt, H.-H. Klauss, and D. Herlach, Phys. Rev. B {\bf 66}, 1444291 (2002).
% Antiferromagnetic ordering of magnetic moments due to cavity trapped electrons in sodium electrosodalite: A zero-field ?SR study

\bibitem{Madsen2004} %bbbbbbb
G. K. H. Madsen, Acta Cryst. A {\bf 60}, 450 (2004).
% The metal-insulator phase transition in mixed potassium±rubidium electro-sodalites

\bibitem{Nakamura2009} %bbbbbbb
K. Nakamura, T. Koretsune, and R. Arita, Phys. Rev. B {\bf 80}, 174420 (2009).
% Ab initio derivation of the low-energy model for alkali-cluster-loaded sodalites

\bibitem{Nakano2010} %aaaaaa
T. Nakano, R. Suehiro, A. Hanazawa, K. Watanabe, I. Watanabe, A. Amato, F. L. Pratt, and Y. Nozue, J. Phys. Soc. Jpn. {\bf 79}, 073707 (2010).
% muSR Study on Antiferromagnetism of Alkali-Metal Clusters Incorporated in Zeolite Sodalite

\bibitem{Nakano2012-SOD} %aaaaaa
T. Nakano, M. Matsuura, A. Hanazawa, K. Hirota, and Y. Nozue, Phys. Rev. Lett. {\bf 109}, 167208 (2012).
% Direct Observation by Neutron Diffraction of Antiferromagnetic Ordering in s Electrons Confined in Regular Nanospace of Sodalite

\bibitem{Nakano2013JKPS}
T.~Nakano, Y.~Ishida, A.~Hanazawa, and Y.~Nozue, J. Korean Phys. Soc. {\bf62}, 2197 (2013).
%Antiferromagnetic Phase Transition of K-Rb Alloy Nanoclusters Incorporated in Sodalite

\bibitem{Nakano2013PRB}
T.~Nakano, H.~Tsugeno, A.~Hanazawa, T.~Kashiwagi, Y.~Nozue, and M.~Hagiwara, Phys. Rev. B {\bf88}, 174401 (2013).
%Antiferromagnetic resonance in alkali-metal clusters in sodalite

\bibitem{Nakano2015Mossbauer} %aaaaaa
T. Nakano, N. Fukuda, M. Seto, Y. Kobayashi, R. Masuda, Y. Yoda, M. Mihara, and Y. Nozue, Phys. Rev. B {\bf 91}, 140101 (2015).
% Synchrotron-radiation-based M?ossbauer spectroscopy of 40K in antiferromagnetic potassium nanoclusters in sodalite


%%% LSX %%% 

\bibitem{Nam2010}
N.~H.~Nam, T.~Ohtsu, T.~Araki, S.~Araki, and Y.~Nozue, J. Phys.: Conf. Ser. {\bf200}, 012062 (2010).
% Magnetic properties of Na-K clusters in low-silica X zeolite doped by pressure loading

\bibitem{Hanh2010}
D.~T.~Hanh, T.~Nakano, and Y.~Nozue, J. Phys. Chem. Solids {\bf71}, 677 (2010).
%Strong dependence of ferrimagnetic properties on Na concentration in Na?K alloy clusters incorporated in low-silica X zeolite

\bibitem{Nozue-Na-LSX2012} 
Y.~Nozue, Y.~Amako, R.~Kawano, T.~Mizukane, and T.~Nakano, J. Phys. Chem. Solids {\bf 73}, 1538 (2012).

\bibitem{Nakano2013K-LSX}
T.~Nakano, D.~T.~Hanh, A.~Owaki, Y.~Nozue, N.~H.~Nam, and S.~Araki, J. Korean Phys. Soc. {\bf63}, 512 (2013).
%Insulator-to-metal Transition and Magnetism of Potassium Metals Loaded into Regular Cages of Zeolite LSX

\bibitem{Igarashi2013-Na-LSX} %aaaaaa
M. Igarashi, T. Nakano, P.T. Thi, Y. Nozue, A. Goto, K. Hashi, S. Ohki, T. Shimizu, A. Krajnc, P. Jegli\v{c}, and D. Ar\v{c}on, Phys. Rev. B {\bf 87}, 075138 (2013).
% NMR study of thermally activated paramagnetism in metallic low-silica X zeolite filled with sodium atoms

\bibitem{Kien2015}
L.~M.~Kien, T.~Goto, D.~T.~Hanh, T.~Nakano, and Y.~Nozue, J. Phys. Soc. Jpn. {\bf84}, 064718 (2015).
%Ferromagnetism of Na?K Alloy Clusters Incorporated in Zeolite Low-Silica X

\bibitem{Igarashi2016}  %aaaaaaa
M. Igarashi, P. Jegli\v{c}, A. Krajnc, R. \v{Z}itko, T. Nakano, Y. Nozue, and D. Ar\v{c}on, Sci. Rep. {\bf 6}, 18682 (2016).
%Metal-to-insulator crossover in alkali doped zeolite

\bibitem{Araki2019}
S.~Araki, N.~H.~Nam, K.~Shimodo, T.~Nakano, and Y.~Nozue, Phys. Rev. B {\bf99}, 094403 (2019).
%Ferromagnetism of Potassium Metal under Pressure-Loading into Zeolite Low-Silica X (LSX)


%LTL
\bibitem{Kelly1995} %aaaaaaa
M. J. Kelly, J. Phys.: Condens. Matter {\bf7}, 5507 (1995).
% A model electronic structure for metal-intercalated zeolites

\bibitem{Anderson1997} %aaaaaaa
P. A. Anderson, A. R. Armstrong, A. Porch, P. P. Edwards, and L. J. Woodall, J. Phys. Chem. {\bf 101}, 9892 (1997).
% Structure and Electronic Properties of Potassium-Loaded Zeolite L


\bibitem{Thi2016} %aaaaaaa
P. T. Thi, T. Nakano, Y. Sakamoto, and Y. Nozue, J. Phys. Soc. Jpn. {\bf85}, 024703 (2016).
% Optical, Electrical and Magnetic Properties of Potassium Metal Loaded into Channel-Type Zeolite L

\bibitem{Thi2017}  %aaaaaaa
P. T. Thi, T. Nakano, Y. Sakamoto, and Y. Nozue, IOP Conf. Ser.: Mater. Sci. Eng. {\bf196}, 012002 (2017).
% Electronic Properties of Alkali Metals Loaded into Channel-Type Zeolite L


%He

\bibitem{Wada2009} %aaaaa
N. Wada, T. Matsushita, M. Hieda, and R. Toda, J. Low Temp. Phys. {\bf 157}, 324 (2009).
% Fluid States of Helium Adsorbed in Nanopores

\bibitem{Yamashita2009} %aaaaaaa
K. Yamashita and D.S. Hirashima, Phys. Rev. B {\bf 79}, 014501 (2009).
% Superfluid density of 4He confined in nanopores

\bibitem{Eggel2011} %aaaaaaa
T. Eggel, M.A. Cazalilla, and M. Oshikawa, Phys. Rev. Lett. {\bf 107}, 275302 (2011).
% Dynamical Theory of Superfluidity in One Dimension

\bibitem{Taniguchi2013} %aaaaaaa
J. Taniguchi, K. Demura, and M. Suzuki, Phys. Rev. B {\bf 88}, 014502 (2013).
% Dynamical superfluid response of 4He confined in a nanometer-size channel

\bibitem{Matsushita2016} %aaaaaaa
T. Matsushita, A. Shinohara, M. Hieda, and N. Wada, J. Low Temp. Phys. {\bf 183}, 273 (2016).
%Superfluid Onset of 4He Nanotube Depending on a One-Dimensional Length

\bibitem{Taniguchi2005} %aaaaaaa
J. Taniguchi, A. Yamaguchi, H. Ishimoto, H. Ikegami, T. Matsushita, N. Wada, S.M. Gatica, M.W. Cole, F. Ancilotto, S. Inagaki, and Y. Fukushima, Phys. Rev. Lett. {\bf 94}, 065301 (2005).
% Possible One-Dimensional 3He Quantum Fluid Formed in Nanopores

%semicon

\bibitem{Bogomolov1978}
V.~N.~Bogomolov, Sov. Phys. Usp. {\bf21}, 77 (1978) [Usp. Fiz. Nauk {\bf124}, 171 (1978)].
% Liquids in ultrathin channels (Filament and cluster crystals)-SovPhysUsp21-Bogomolov1978

\bibitem{Nozue-Se-1990} %aaaaaaa
Y. Nozue, T. Kodaira, O. Terasaki, K. Yamazaki, T. Goto, D. Watanabe, and J.M. Thomas, J. Phys.: Condens. Matter {\bf 2}, 5209 (1990).
% Absorption spectra of selenium clusters and chains incorporated into zeolites

\bibitem{Tang1991} %aaaaaaa
Z. K. Tang, Y. Nozue, and T. Goto, J. Phys. Soc. Jpn. {\bf 60}, 2090 (1991).
% Quantum Size Effect on the Excitation Energy and the Oscillator Strength of Pbl2 Clusters in Zeolite

\bibitem{Tang1992} %aaaaaaa
Z. K. Tang, Y. Nozue, and T. Goto, J. Phys. Soc. Jpn. {\bf 61}, 2943 (1992).
% Quantum Size Effect on the Excited State of Hgl2 , Pbl2 and BiI3 Clusters and Molecules in Zeolite LTA

\bibitem{Demkov1995} %ref3
A. A. Demkov, and O.F. Sankey, {\it Access in Nanoporous Materials}, Eds. T. J. Pinnavaia and M. F. Thorpe  (Plenum Press, New York, 1995), p. 27.
% “Electronic structure theory for zeolites” (invited review)

\bibitem{Demkov1996a} %ref3
A. A. Demkov, and O.F. Sankey,  J. Comp. Aided Mater. Design {\bf 3}, 128 (1996).
%“Clusters stuffed inside frameworks: electronic structure theory,”

\bibitem{Demkov1996b} %ref3
A. A. Demkov, and O.F. Sankey,  Chem. Mater. {\bf 8}, 1793 (1996).
% “Recent developments in the theory of supralattices (review),”

\bibitem{Demkov1997} %ref3
A. A. Demkov, and O.F. Sankey, Phys. Rev. B {\bf 56}, 10497 (1997).
% “Model simulations of Supralattices: semiconductor Si clusters in Zeolites',”

\bibitem{Demkov2001} %ref3
A. A. Demkov and O. F. Sankey,  J. Phys.: Condens. Matter {\bf 13}, 10433 (2001).
% “Theory of zeolite supralattices: Se in zeolite LTA,”

\bibitem{Goldbach2005} %ref
A. Goldbach and M.-L Sabounghi, Acc. Chem. Res. {\bf 38}, 705 (2005).
%“Selenium/Zeolite Y Nanocomposites,”


\bibitem{Heer1987}
W.~A.~de~Heer, W.D. Knight, M.Y. Chou and M.L. Cohen, Solid State Phys. {\bf40}, 93 (1987).
% Electronic Shell Structure and Metal Clusters


\bibitem{Tachiki1968}
M.~Tachiki, J. Phys. Soc. Jpn. {\bf25}, 686 (1968).
% Exchange Contribution to the D-Parameter of Mn2+ in CoCl2-2H2O


\bibitem{Holm1967} 
R.~Holm, {\it Electric Contacts, Theory and Applications}, 4th ed. Springer, New York, 1967.

\bibitem{Kelemen1992} 
G.~Kelemen and G.~Schon, J. Mater. Sci. {\bf27}, 6036 (1992).

\bibitem{Pincus1960}
P. Pincus, Phys. Rev. Lett. {\bf 5}, 13 (1960).
% a-hematite (a-Fe2O3)

\bibitem{Ziese1998}
M.~Ziese and C.~Srinitiwarawong, Phys. Rev. B {\bf 58}, 11519 (1998).
% Polaronic effects on the resistivity of manganite thin films

\bibitem{Shinozuka1987}
Y.~Shinozuka, J. Phys. Soc. Jpn. {\bf56}, 4477 (1987).  
% Self-Consistent Mean-Field Theory for Interacting Electrons in Three-Dimensional Deformable Lattice

\bibitem{Anderson1975}
P.~W. Anderson, Phys. Rev. Lett. {\bf34}, 953 (1975).
% Model for the Electronic Structure of Amorphous Semiconductors

\bibitem{Toyozawa1981}
Y.~Toyozawa, J. Phys. Soc. Jpn. {\bf50}, 1861 (1981).
% Charge Tran sf er Instability with Structural Change. I. Two-Sites Two-Electrons System

\bibitem{Moriya1980}
T.~Moriya and H.~Hasegawa, J. Phys. Soc. Jpn. {\bf48}, 1490 (1980).
% A Unified Theory of Magnetism in Narrow Band Electron Systems

%\bibitem{Nakamura2009}
%K.~Nakamura, T.~Koretsune, and R.~Arita, Phys. Rev. B {\bf80}, 174420 (2009).


\end{thebibliography}

%%%%%%%%%%%%%%%%%%%%%%%%%%%%%%%%%%%%%

\end{document}